\documentclass[graybox]{svmult}

\usepackage[english]{babel}
\usepackage[utf8x]{inputenc}
\usepackage[T1]{fontenc}

\usepackage[a4paper,top=3cm,bottom=2cm,left=3cm,right=3cm,marginparwidth=1.75cm]{geometry}

\usepackage{amsmath}
\usepackage{amsfonts}
\usepackage{amssymb}
\usepackage{amsbsy}
\usepackage{bm}
\usepackage{trfsigns}
\usepackage{mathptmx}       
\usepackage{helvet}         
\usepackage{courier}        
\usepackage{type1cm}        

\usepackage{makeidx}         
\usepackage{graphicx}        
\usepackage{multicol}        
\usepackage[bottom]{footmisc}

\usepackage[]{algorithm2e}
\usepackage{graphicx}
\usepackage{xcolor}
\usepackage[colorinlistoftodos]{todonotes}
\usepackage[colorlinks=true, allcolors=blue]{hyperref}
\usepackage{tikz}
\def\checkmark{\tikz\fill[scale=0.4](0,.35) -- (.25,0) -- (1,.7) -- (.25,.15) -- cycle;}
\usepackage{slashbox}
\usepackage[square,numbers,sort]{natbib}
\newcommand{\R}{\mathbb{R}}

\newcommand{\dims}[1]{{n_{#1}}}
\newcommand{\makered}[1]{\color{black}#1\color{black}}
\makeindex             

\begin{document}
\title*{Scalable Inference of Ordinary Differential Equation Models of Biochemical Processes}
\author{Fabian Fr\"ohlich, Carolin Loos, Jan Hasenauer}
\institute{
{\small \normalfont \textit{To appear in the book "Gene Regulatory Networks: Methods and Protocols"}}\\~\\
Fabian Fr\"ohlich \at Institute of Computational Biology, Helmholtz Zentrum M\"unchen, 85764 Neuherberg, Germany and Chair of Mathematical Modeling of Biological Systems, Center for Mathematics, Technische Universit\"at
M\"unchen, 85748 Garching, Germany, \email{fabian.froehlich@helmholtz-muenchen.de}
\and Carolin Loos \at Institute of Computational Biology, Helmholtz Zentrum M\"unchen, 85764 Neuherberg, Germany and Chair of Mathematical Modeling of Biological Systems, Center for Mathematics, Technische Universit\"at
M\"unchen, 85748 Garching, Germany, \email{carolin.loos@helmholtz-muenchen.de}
\and Jan Hasenauer \at Institute of Computational Biology, Helmholtz Zentrum M\"unchen, 85764 Neuherberg, Germany and Chair of Mathematical Modeling of Biological Systems, Center for Mathematics, Technische Universit\"at
M\"unchen, 85748 Garching, Germany \email{jan.hasenauer@helmholtz-muenchen.de}}
\maketitle

\abstract{
Ordinary differential equation models have become a standard tool for the mechanistic description of biochemical processes. If parameters are inferred from experimental data, such mechanistic models can provide accurate predictions about the behavior of latent variables or the process under new experimental conditions. Complementarily, inference of model structure can be used to identify the most plausible model structure from a set of candidates, and, thus, gain novel biological insight.
Several toolboxes can infer model parameters and structure for small- to medium-scale mechanistic models out of the box. However, models for highly multiplexed datasets can require hundreds to thousands of state variables and parameters. For the analysis of such large-scale models, most algorithms require intractably high computation times. This chapter provides an overview of state-of-the-art methods for parameter and model inference, with an emphasis on scalability.}

\section{Introduction}

In systems biology, ordinary differential equation (ODE) models have become a standard tool for the analysis of biochemical reaction networks~\cite{KlippBook2005}. The ODE models can be derived from information about  the underlying biochemical processes~\cite{Kitano2002,Kitano2002b} and allow the systematic integration of prior knowledge. ODE models are particularly valuable as they can be used to predict the temporal evolution of latent variables~\cite{AdlungKar2017,BuchholzFlo2013}. Moreover, they provide executable formulations of biological hypotheses and therefore allow the rigorous falsification of hypotheses~\cite{IntosalmiNou2016,HugSch2016,HrossFie2016,ToniOza2012,MolinelliKor2013,SchillingMai2009}, thereby deepening the biological understanding. Furthermore, ODE models have been applied to derive model-based biomarkers~\cite{FeyHal2015,EduatiDol2017,HassMas2017}, that enable a personalized design of targeted therapies in precision medicine.

To construct predictive models, model parameters have to be inferred from experimental data. This inference requires the repeated numerical simulation of the model. Consequently, parameter inference is computationally demanding if the required computation time for the numerical solution is high. For many applications, small- and medium-scale models, i.e., models consisting of a small number of species belonging to the core pathway, accurate prediction and hypothesis testing~\cite{MaiwaldHas2016,Kitano2002,SnowdenGra2017,TranstrumQiu2016}. Low-dimensional models can be derived directly or obtained from large-scale models by model reduction, e.g., by lumping multi-step reactions to one-step reactions~\cite{DanoMad2006} or by assuming time-scale separation~\cite{Klonowski1983}. These small- and medium-scale models can be analyzed using established toolboxes implementing state-of-the-art methods~\cite{HoopsSah2006,RaueSte2015,SomogyiBou2015}.

For models describing few conditions, e.g., the response of a single cell line to a small set of stimulations, the lumping and ignoring of processes might be appropriate. Yet, if a model is to be used for a wide range of conditions, e.g., to describe the responses of many cell lines to many different stimuli, a detailed mechanistic description is required~\cite{SantosVer2007,YaoPil2016,AdlungKar2017} as simplifications only hold for selected conditions.
Detailed mechanistic, generalizing models appear particularly valuable for  precision medicine, where the model must accurately predict treatment outcomes for many different patients~\cite{OgilvieKov2017}. These comprehensive models, typically describing most species in multiple different pathways including respective crosstalk, can easily describe thousands of molecular species involved in thousands of reactions with thousands of parameters. For such models, parameter inference is often intractable as it is prohibitively computationally expensive~\cite{SchillingsSun2015,BabtieStu2017}.

Beyond model parameters, also the model structure might be unknown, e.g., the biochemical reactions or their regulations might be unknown~\cite{OconeHag2015,KondoferskyFuc2015}. Then, inference of model structure can be used to generate new mechanistic insights. For ODE models this can be achieved by constructing multiple model candidates corresponding to different biological hypotheses. These hypotheses can be falsified using model selection criteria such as the Akaike Information Criterion (AIC)~\cite{Akaike1973}, or Bayesian Information Criterion (BIC)~\cite{Schwarz1978}. For large models, a high number of mutually non-exclusive hypotheses is not uncommon and typically leads to a combinatorial explosion of the number of model candidates~(see, e.g., \cite{SteiertTim2016,KlimovskaiaGan2016,LoosMoe2017}). Computing the AIC or BIC for all model candidates for comparison would require parameter inference for each model candidate and may seem futile, given that parameter inference for a single model can already be challenging. 

In this chapter, we will review scalable methods that render model parameter and model structure inference tractable for large-scale ODE models, which have hundreds to thousands of molecular species, biochemical reactions and parameters. For parameter inference, we will focus on different gradient-based optimization schemes and describe their scaling properties with respect to the number of molecular species and number of parameters. For inference of model structure, we will focus on complexity penalization schemes that allow the simultaneous inference of model structure and parameters and thus scale better than linearly with the number of model candidates.

\section{Inference of Model Parameters}

An ODE model describes the temporal evolution of the concentrations of $\dims{x}$ different molecular species $x_i$. The dynamics of $\textbf{x}$ are determined by the vector field $f$ and the initial condition $\textbf{x}_0$:
\begin{equation}
\dot{\textbf{x}}=f(t,\textbf{x},\boldsymbol{\theta}),\quad \textbf{x}(t_0) = \textbf{x}_0(\boldsymbol{\theta}).
\label{eq: ODE}
\end{equation}
Both of these functions may depend on the unknown parameters $\boldsymbol{\theta}\in\Theta\subset\R^{\dims{\theta}}$ such as kinetic rates. The parameter domain $\Theta$ can constrain the parameter values to biologically reasonable numbers.

In general, $\textbf{x}$ and $f$ can also be derived from a discretization of a partial differential equation model~\cite{HockHas2013,HrossPhDThesis2017,MenshykauGer2013,HrossHas2016} or describe the temporal evolution of empirical moments of stochastic processes~\cite{FroehlichTho2016,Ruess2015,MunskyTri2009}. 

Experiments usually provide information about $\dims{y}$ different observables $y_i$ which depend linearly or nonlinearly on the concentrations $\textbf{x}$. A direct measurement of $\textbf{x}$ is usually not possible. The dependence of the observable on concentrations and parameters is described by 
\begin{equation}
\textbf{y}(t,\boldsymbol{\theta}) = h(\textbf{x}(t,\boldsymbol{\theta}),\boldsymbol{\theta}).
\label{eq: observable}
\end{equation}

\subsection{Problem Formulation}

To build predictive models, the model parameters $\boldsymbol{\theta}$ have to be inferred from experimental data. This inference problem is usually formulated as an optimization problem. In this optimization problem, an objective function $J(\boldsymbol{\theta})$, describing the difference between measurements and simulation, is minimized. In the following, we will first formulate the optimization problem and then discuss methods to solve it efficiently.

Experimental data are subject to measurement noise. A common assumption is that the measurement noise for all time-points $t_j$ and observables $y_i$ is additive and independent, normally distributed for all time-points:
\begin{equation}
\bar{y}_{ij} = y_i(t_j,\boldsymbol{\theta}) + \epsilon_{ij}, \quad \epsilon_{ij} \overset{id}{\sim} \mathcal{N}(0,\sigma_{ij}^2(\boldsymbol{\theta})).
\end{equation}
At each of the $T$ time-points $t_j$ up to $\dims{y}$ different measurements $\bar{y}_{ij}$ can be recorded in the experimental data $\mathcal{D} = \{((\bar{y}_{ij})_{i=1}^{\dims{y}},t_j)\}_{j=1}^{T}$.
As the standard deviation of the measurement noise is potentially unknown, we model it as $\sigma_{ij}(\boldsymbol{\theta})$. This yields the likelihood function
\begin{equation}
p(\mathcal{D}|\boldsymbol{\theta}) = \prod_{j=1}^T\prod_{i=1}^\dims{y}\frac{1}{\sqrt{2\pi\sigma_{ij}^2(\boldsymbol{\theta})}}\mathrm{exp}\left(-\frac{1}{2}\left(\frac{\bar{y}_{ij}-y_i(t_j,\boldsymbol{\theta})}{\sigma_{ij}(\boldsymbol{\theta})}\right)^2\right).
\label{eq: likelihood}
\end{equation}
Other plausible noise assumptions include log-normal distributions, which correspond to multiplicative measurement noise~\cite{Kreutz2007}. Distributions with heavier tails, such as the Laplace distribution, can be used to increase robustness to outliers in the data~\cite{MaierLoo2017}.

The model can be inferred from experimental data by maximizing the likelihood~\eqref{eq: likelihood}, which yields the maximum likelihood estimate (MLE). However, the evaluation of the likelihood function, $p(\mathcal{D}|\boldsymbol{\theta})$, involves the computation of several products, which can be numerically unstable. Thus, the negative log-likelihood 
\begin{equation}
\begin{aligned}
J(\boldsymbol{\theta}) = -\log(p(\mathcal{D}|\boldsymbol{\theta})) =&  \frac{1}{2} \sum_{i=1}^\dims{y} \sum_{j=1}^T \log\left(2\pi\sigma_{ij}^2(\boldsymbol{\theta})\right) + \left(\frac{\bar{y}_{ij} - y_{i}(t_j,\boldsymbol{\theta})}{\sigma_{ij}(\boldsymbol{\theta})}\right)^2
\label{eq: objective function}
\end{aligned}
\end{equation}
is often used as objective function for minimization. As the logarithm is a strictly monotonously increasing function, the minimization of $ J(\boldsymbol{\theta})=-\log(p(\mathcal{D}|\boldsymbol{\theta}))$ is equivalent to the maximization of $ p(\mathcal{D}|\boldsymbol{\theta})$. Therefore, the corresponding minimization problem
\begin{equation}
\boldsymbol{\theta}^* = \arg \min_{\boldsymbol{\theta} \in \Theta} J(\boldsymbol{\theta}),
\label{eq: optimization problem}
\end{equation}
 will infer the MLE parameters. If the noise variance $\sigma_{ij}^2$ does not depend on the parameters $\boldsymbol{\theta}$, \eqref{eq: objective function} is a weighted least-squares objective function. As we will discuss later, this least-squares structure can be exploited by several optimization methods.

If prior knowledge about the parameters is available, this can be encoded in a prior probability $p(\boldsymbol{\theta})$. According to Bayes' theorem~\cite{PugaKry2015}, the posterior probability $p(\boldsymbol{\theta}|\mathcal{D})$ is defined by:
\begin{equation}
p(\boldsymbol{\theta}|\mathcal{D}) = \frac{p(\mathcal{D}|\boldsymbol{\theta})p(\boldsymbol{\theta})}{p(\mathcal{D})}.
\label{eq: poster}
\end{equation}
The evidence $p(\mathcal{D})$ is usually difficult to compute. However, as it is independent of $\boldsymbol{\theta}$, the respective term can be omitted for parameter inference. This yields the objective function
\begin{equation}
J(\boldsymbol{\theta}) = -\log(p(\mathcal{D}|\boldsymbol{\theta})) - \log(p(\boldsymbol{\theta})),
\label{eq: posterior}
\end{equation}
which corresponds to the log-posterior up to the constant $\log(p(\mathcal{D}))$. The respective optimization problem yields the maximum a posteriori estimate (MAP). 

\begin{figure}[tb]
\centering
\includegraphics[scale=0.85]{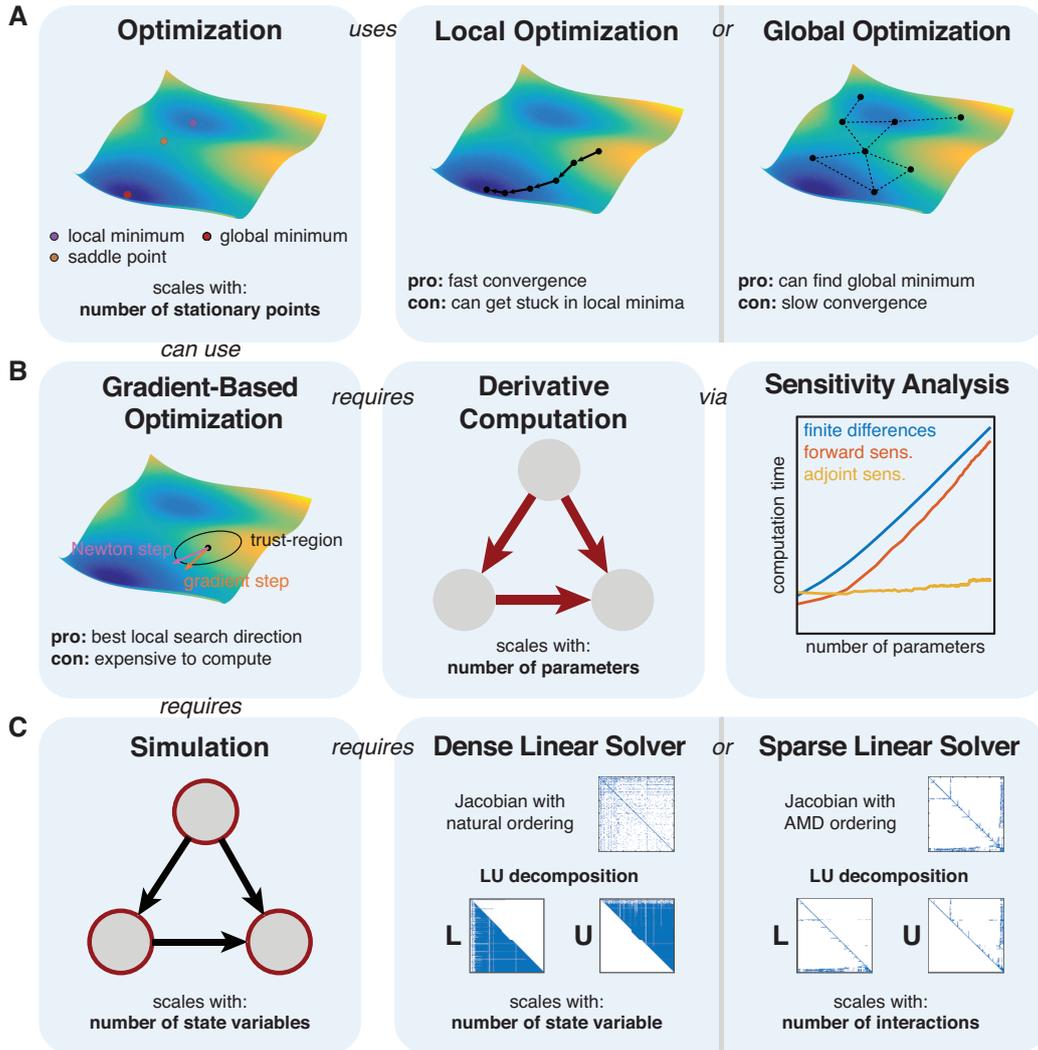}
\caption{Overview of numerical methods and scaling properties for parameter inference. (A)~Schematic scaling properties of optimization. Icons and properties of local and global optimizations are shown on the right. (B)~Examples of gradient-based methods to determine parameter updates. Computation times of different approaches for gradient computation are shown on the right. (C)~Schematic scaling properties of simulation algorithms. Dense and sparse direct linear solvers are compared on the right. Reordering was performed using the Approximative Minimum Degree (AMD) ordering algorithm~\cite{AmestoyDav1996}}\label{fig:optimization}
\end{figure}

\subsubsection{Properties of the Optimization Problem}

The optimization problem \eqref{eq: objective function} is convex in $y_{i}(t_j,\boldsymbol{\theta})$, but usually non-convex in $\boldsymbol{\theta}$. Thus, the objective function $J(\boldsymbol{\theta})$ can possess local minima and saddle points. Local minima can be problematic as optimization algorithms may get stuck, yielding a suboptimal agreement between experimental data and model simulation. Interestingly, recent literature suggests that saddle points might affect the efficiency of optimization  more severely than local minima~\cite{DauphinPas2014}. For unconstrained problems, saddle points and local minima are both stationary points $\boldsymbol{\theta}^{*}$ at which the gradient vanishes
\begin{equation}
\nabla J(\boldsymbol{\theta}^{*}) = 0,
\end{equation}
i.e., they both satisfy a necessary local optimality condition (see Figure~\ref{fig:optimization}A left for an example). The sufficient condition for a local minimum is the positive definiteness of the Hessian $\nabla^2 J(\boldsymbol{\theta}^{*})$, which indicates that there are only directions of positive curvature. For a saddle point the Hessian $\nabla^2 J(\boldsymbol{\theta}^{*})$ is indefinite or semi-definite, which indicates that there may be directions of negative or zero curvature.

The dependence of the number of local minima and saddle points for ODE models on the number of parameters is poorly understood. For deep learning problems, an exponential increase in the number of local minima with the number of parameters is primarily attributed to parameter permutation symmetries~\cite{AnandkumarGe2016}, which are rare in ODE models. Yet, for deep learning problems, saddle points are also problematic as they affect the performance of local optimization methods~\cite{DauphinPas2014}. Arguments for an exponential increase in stationary points with the number of parameters are often based on random matrix theory~\cite{DauphinPas2014} and rely on strong assumptions on the distribution of entries in the Hessian of the objective function $J(\boldsymbol{\theta})$. These assumptions have to be rigorously checked as they can lead to wrong conclusions, as shown for the stability of ODEs~\cite{KirkRol2015}. As the objective function $J(\boldsymbol{\theta})$ depends on the solution to an ODE, the validity of such assumptions is not evident and difficult to assess rigorously. For saddle points, we are not aware of any rigorous evaluation. Thus, the exact dependence of the number of local minima and saddle points on the parameter dimensionality remains elusive.

\subsection{Optimization Methods} \label{Optimization Methods}

The infamous No Free Lunch Theorem for optimization~\cite{WolpertMac1997} states that there exists no single optimization method that performs best on all classes of optimization problems. Accordingly, empirical evidence as well as a careful analysis of the problem structure should be considered when selecting a suitable optimization method. \makered{ The plethora of different optimization methods commonly used in systems biology can be classified as 
\begin{enumerate}
\item \textbf{local} and \textbf{global} methods,
as well as
\item \textbf{gradient-based} and \textbf{derivative-free} methods.
\end{enumerate}
Local methods search for local optima, while global methods search for global minima. The separation into local and global methods is often not clear-cut. Thus, methods, such as simulated annealing, are sometimes classified as local and sometimes as global methods~\cite{GoffeFer1994,Neumaier2004,JohnsonMcG1997}. Therefore, the following paragraph contains many soft statements that should only serve as guidelines. Gradient-based methods exploit first and potentially higher order derivatives of the objective function, while derivative-free methods solely use the objective function.}

\begin{table}[t]
\begin{center}
\caption{Examples for local and global, as well as derivative-free and gradient-based optimization algorithms }
\label{tab: classification optimization methods}
\begin{tabular}{|c||c|c|}
\hline
      &  Derivative-Free & Gradient-Based   \\ 
\hline
\hline
      &  Pattern Search~\cite{HookeJee1961} & Gradient Descent~\cite{NocedalWri2006}   \\ 
Local &  Nelder-Mead~\cite{NelderMea1965} & Newtons Method~\cite{NocedalWri2006}   \\ 
      &  Hill Climbing~\cite{DeLaMazaYur1994} & Levenberg-Marquardt~\cite{Levenberg1944,Marquardt1963}   \\ 
\hline
       & Genetic Algorithm~\cite{Holland1992} & Multi-Start~\cite{Neumaier2004}   \\ 
Global & Particle Swarm~\cite{Kennedy2011} & Scatter Search~\cite{Egea2007}   \\ 
       & Simulated Annealing~\cite{KirkpatrickGel1983} & Clustering Search~\cite{KanTim1987}    \\ 
\hline
\end{tabular}
\end{center}
\end{table}

Local methods construct a sequence of trial points that successively decrease the objective function values (See Figure~\ref{fig:optimization}A middle). This procedure is usually faster than global methods, but can get stuck in local minima~\cite{AshyraliyevFom2009}. Most local derivative-free methods are  direct search methods~\cite{HookeJee1961}. In contrast to local methods, global methods often rely on a population of trial points which are iteratively refined (see Figure ~\ref{fig:optimization}A right). This can increase the chance of reaching the global minimum, but usually slower~\cite{AshyraliyevFom2009}. Global derivative-free methods mostly employ stochastic schemes, which are often inspired by nature~\cite{FisterYan2013}, while global gradient-based methods usually perform repeated local optimizations. Examples of local and global as well as respective derivative-free and gradient based methods are given in Table~\ref{tab: classification optimization methods}.

Not all global methods are guaranteed to converge to the global minimum~\cite{Neumaier2004}. Convergence to the global minimum is only guaranteed for rigorous and (asymptotically) complete global methods, such as branch-and-bound~\cite{LawlerWoo1966} and grid search~\cite{Neumaier2004}. As long as only local information, i.e., function values and respective parameter derivatives, is available, the termination of these methods will require exponentially expensive dense search~\cite{Neumaier2004,TornZil1989}. The termination of global methods is crucial, as it might be relatively easy to find the global minimum but comparably hard to guarantee that  it is indeed the global minimum. Global information, such as Lipschitz constants, is rarely available for ODE problems, such that dense, i.e., exhaustive, search would be necessary for guaranteed convergence. As the parameters $\boldsymbol{\theta}$ are generally continuous, dense search is rarely possible. Instead, meta-heuristics for termination and optimization are employed. 

For many meta-heuristic methods, there exists little to no theoretical justification or convergence proofs~\cite{Neumaier2004}. Others may converge with probability arbitrarily close to 1, but might only do so after infinitely many function evaluations. In practice, many meta-heuristic algorithms even fail to work reliably for smooth, convex problems with few parameters~\cite{RiosSah2013}. In fact, for some algorithms, non-convergence can even be proven mathematically~\cite{Rudolph1994}. Eventually, even a rigorous convergence guarantee will be useless if the convergence rate is too slow for practical purposes. For most methods, there exists a plethora of disparate variants, which renders comprehensive analysis of convergence proofs and convergence rates challenging. This is quite unsatisfying from a theoretical perspective. In practice, reasonable results can be obtained using global optimization methods~\cite{MolesMen2003,RaueSch2013,Banga2008}. Yet, usually no guarantees of global optimality can be given.

For the remainder of this chapter we will primarily focus on global gradient-based optimization methods, which typically rely on repeated local optimization. For these meta-heuristic methods, a local optimization is started at (random) points in parameter space. The termination of these methods usually relies on a specified maximum number of local optimizations, but also Bayesian methods can be applied~\cite{BoenderKan1987}. For the convergence to the global minimum, the local optimization has to be started in the region of attraction of the global optimum. Thus, the probability of convergence will depend on the relative volume of the region of attraction with respect to the search domain $\boldsymbol{\theta}$. Several adaptive methods, such as scatter search~\cite{Egea2007} or clustering search~\cite{KanTim1987}, try to improve the chance of starting a local optimization in the region of attraction of the global optimum, but rely on the embeddedness of the global optimum~\cite{TornAli1999}. The embeddedness of the global optimum characterizes how well local minima cluster and determines how indicative the objective function value at the starting point is of the chance of converging to the global minimum. We are not aware of any analysis of embeddedness of the global minimum for models in systems biology, but it is likely to be problem dependent. The resulting rate of convergence will be determined by the rate of convergence of the local method and the probability to sample a starting point from the region of attraction of the global optimum. 

For most methods that employ repeated local optimization, the individual local optimization runs can trivially be run in parallel~\cite{FroehlichKal2017}, which enables efficient use of high performance computing structure. Moreover, multiple global runs can be asynchronously parallelized to enhance efficiency through cooperativity~\cite{PenasGon2015}. Following recent studies~\cite{RaueSch2013,HrossHas2016}, we deem this repeated local optimization a suitable candidate for scalable optimization and will in the following discuss the properties of respective local gradient based methods (see Figure~\ref{fig:optimization}B) in more detail.

\subsubsection{Line-Search Methods}

Line-search methods are local optimization methods that iteratively update the parameter values in direction $\textbf{s}\in\mathbb{R}^{\dims{\theta}}$, such that the objective function value $J(\boldsymbol{\theta})$ is successively reduced: 
\begin{equation}
\boldsymbol{\theta}_k = \boldsymbol{\theta}_{k-1}+\gamma \cdot \textbf{s}\quad s.t.\quad J(\boldsymbol{\theta}_k)<J(\boldsymbol{\theta}_{k-1}),\gamma>0,
\end{equation}
where $\gamma\in\mathbb{R_+}$ controls the step size. For line-search, the update direction $\textbf{s}$ is fixed first and then a suitable $\gamma$ is determined. The alternative to line-search methods are trust-region methods which first define a maximum step length and then find a suitable update direction. We will discuss trust-region methods in more detail in the following subsection. The classification in line-search or trust-region methods can be ambiguous and may depend on the specific implementation of the method. This will be discussed in more detail at the end of this subsection. In this chapter, we will follow the classification of Nocedal and Wright, who also provide an excellent discussion of the topic~\cite{NocedalWri2006}.

Line-search methods are particularly appealing as they reduce the possibly high-dimensional optimization problem to a sequence of one-dimensional problems of finding good values for $\gamma$. To ensure convergence, $\gamma$ has to meet certain conditions~\cite{Armijo1966,Wolfe1969}. Methods to determine the step size that satisfy these conditions are sometimes referred to as globalization techniques. Note that enforcing these conditions only guarantees convergence to a local stationary point, i.e., a local minimum, local maximum or saddle point, but not to a global minimum~\cite{NocedalWri2006}.

It is more or less well established that for local optimization, gradient-based methods should be used when applicable. Kolda et \textit{al.} state in their review on direct search methods that  "Today, most people’s first recommendation (including ours) to solve an unconstrained problem for which accurate first derivatives can be obtained would not be a direct search method, but rather a gradient-based method"~\cite{KoldaLew2003}. Lewis et \textit{al.} claim that "the success of quasi-Newton methods, when applicable, is now undisputed"~\cite{LewisTor2000}. Gradient-based methods, such as quasi-Newton methods, are applicable when the gradient (and the Hessian) of the objective function is continuous and can be computed accurately. By definition, the gradient $\nabla J(\boldsymbol{\theta})$ is only continuous with respect to parameters when the objective function is continuously differentiable. Higher order continuous differentiability corresponds to continuity of the respective higher order derivatives, such as the Hessian. For many noise distributions, such as the normal and log-normal distribution, the negative log-likelihood is infinitely often continuously differentiable with respect to the observables, given that a finite number of measurements is considered. Thus, the continuity of derivatives of the objective function \eqref{eq: objective function} only depends on the continuity of derivatives of the model output \eqref{eq: observable}. However, for particular noise distributions, such as the Laplace distribution, the negative log-likelihood may not be differentiable with respect to the model outputs. In the following, we will assume that both, \eqref{eq: observable} and \eqref{eq: objective function}, are twice continuously differentiable with respect to the parameters.

For continuously differentiable objective functions, an intuitive choice for $\textbf{s}$ is the gradient
\begin{equation}
\textbf{s}_{\mathrm{grad}}=-\nabla J(\boldsymbol{\theta}_{k-1}).
\end{equation}
Optimization methods using this search direction are called gradient descent methods. Locally, the gradient provides the steepest descent with respect to the euclidean norm. This means that the gradient points in the direction $\textbf{d}$, with unit length in the euclidean norm, which yields the strongest decrease of $J(\boldsymbol{\theta})$ in a neighborhood around $\boldsymbol{\theta}$:
\begin{equation}
\frac{\nabla J(\boldsymbol{\theta})}{\|\nabla J(\boldsymbol{\theta})\|_2}=\underset{\textbf{d}:\|\textbf{d}\|_2=1}{\mathrm{argmin}}\nabla J(\boldsymbol{\theta})^T\textbf{d}.
\end{equation}
Yet, depending on the objective function, this neighborhood might be arbitrarily small, resulting in small values $\gamma$. This, for instance, is the case for objective functions with curved ridges, e.g., the Rosenbrock function~\cite{Rosenbrock1960}, which can arise from (non-linear) dependencies between parameters. Moreover, gradient descent methods take small steps in the vicinity of saddle points~\cite{DauphinPas2014}, which can lead to high iteration numbers or premature termination in individual optimization runs.

\makered{The issue of small step sizes is addressed in the Newton's method by including the Hessian $\nabla^2 J(\boldsymbol{\theta}_{k-1})$, which encodes the curvature, in the calculation of the search direction $\textbf{s}$}:
\begin{equation}
 \textbf{s}_{\mathrm{newt}}=- \nabla^2 J(\boldsymbol{\theta}_{k-1}) ^{-1}\nabla J(\boldsymbol{\theta}_{k-1}).
\label{eq:Newton step}
\end{equation}
\makered{For the classical Newton's method, the step size is fixed to $\gamma=1$. However, many modern implementations implement Newton's method as line-search using 1 as default value and adapting the step size if necessary~\cite{NocedalWri2006}}. 

As the Hessian is symmetric, highly efficient methods such as Cholesky factorization can be used to solve this problem for convex optimization problems. However, for ODE models, the computation of the Hessian itself will usually be computationally far more expensive than the computation of the Newton step. Thus, the computational cost of solving \eqref{eq:Newton step} is usually negligible for objective functions depending on ODE models.

\makered{For non-convex problems, the computation of the Newton step ~\eqref{eq:Newton step} may be an ill-posed or not even well-defined problem~\cite{NocedalWri2006}. Moreover, the Newton step might not be a descent direction. The Newton step is only a direction of descent if the scalar product with the gradient is negative:}
\begin{equation}
\textbf{s}_{\mathrm{newt}}^T\cdot \nabla J(\boldsymbol{\theta}_{k-1})<0.
\end{equation}
By substituting the formula for the Newton step, we obtain the following inequality:
\begin{equation}
-\nabla J(\boldsymbol{\theta}_{k-1})^{T} \left(\nabla^2 J(\boldsymbol{\theta}_{k-1})\right)^{-1} \nabla J(\boldsymbol{\theta}_{k-1})<0,
\end{equation}
which is globally satisfied only if the inverse of the Hessian $\nabla^2 J(\boldsymbol{\theta}_{k-1})^{-1}$ is positive definite, i.e., the problem \eqref{eq: optimization problem} is convex. As previously discussed, \eqref{eq: optimization problem} is typically non-convex, thus simple Newton steps will not always yield a direction of descent. Moreover, in the vicinity of a saddle point, the Newton step may point in the direction of the saddle point, thus attracting the optimizer to saddle points~\cite{Yuan2015,DauphinPas2014}.

In the literature, several modifications of Newton's method that always yield descent directions have been proposed~\cite{Levenberg1944,Marquardt1963,Hartley1961}. The Gauss-Newton~\cite{Hartley1961} method exploits the least-squares structure of the objective function \eqref{eq: objective function} and constructs a positive semi-definite approximation to $\nabla^2 J(\boldsymbol{\theta}_{k-1})$. Levenberg~\cite{Levenberg1944} and Marquardt~\cite{Marquardt1963} independently extended this method by introducing a dampening term in the step equation. This yields the Levenberg-Marquardt method
\begin{equation}
-(\boldsymbol{\hat{H}}(\boldsymbol{\theta}_{k-1}) +\lambda \boldsymbol{I})\textbf{s}=\nabla J(\boldsymbol{\theta}_{k-1}),
\label{eq:LM step}
\end{equation}
where $\boldsymbol{\hat{H}}$ is the positive semi-definite Gauss-Newton approximation to the Hessian, $\lambda\geq0$ is the dampening factor and $\boldsymbol{I}$ is the identity matrix. \makered{The magnitude of the dampening factor $\lambda$ regulates the conditioning of \eqref{eq:LM step}. The geometric interpretation of $\lambda$ is that it allows an interpolation between a gradient and a Gauss-Newton step, where $\lambda=0$ corresponds to a pure approximate Newton step}.  Due to the positive-definiteness of the Gauss-Newton approximation, the respective methods cannot follow directions of negative curvature and are thus not attracted to saddle points, but again limited to small step-sizes in the vicinity of saddle points~\cite{DauphinPas2014}.

As the Gauss-Newton method is limited to least-squares problems, the traditional formulation of the Levenberg-Marquardt method has the same limitation. However, it is possible to apply the dampening of the Hessian without the Gauss-Newton approximation. The resulting algorithms are often still referred to as Levenberg-Marquardt method~\cite{NesterovPol2006}. In such a setting, the dampening is often chosen according to the smallest negative eigenvalue~\cite{NocedalWri2006}, using e.g., the Lanczos method~\cite{Lanczos1950}, to ensure the construction of a direction of descent. In the vicinity of saddle points, these methods can be modified to also follow directions of negative curvature~\cite{Nash1984}.

\makered{An alternative to determine the dampening factor $\lambda$ is the use of a trust-region method. A trust region method fixes $\|\gamma \boldsymbol{s}\|=\Delta$ and then determines an approximately matching $\lambda$~\cite{NocedalWri2006}. Thus, Levenberg-Marquardt algorithms can be implemented as line-search methods or as trust-region methods, depending on how $\lambda$ and $\gamma$ are computed. In the following we will discuss trust-region methods more generally.}

\subsubsection{Trust-Region Methods}

\makered{Line-search methods determine a search direction $d$ first and then identify a good step size $\gamma$. Trust region methods do the converse, by specifying a maximum step-size first and then identifing a good search direction~\cite{ByrdSchnabel1987,Sorensen1982,NocedalWri2006}. This allows trust region methods to make large steps close to saddle points and always yield descent directions.} Within the trust-region, $J(\boldsymbol{\theta})$ is replaced by a local approximation, giving rise to the trust-region subproblem. Most trust-region algorithms use the objective function derivatives to construct a quadratic trust-region subproblem
\begin{equation}
\underset{\textbf{s}\in B_{\textbf{D},\Delta}(\boldsymbol{\theta})}{\min}\,\frac{1}{2}\textbf{s}^T\nabla^2J(\boldsymbol{\theta})\textbf{s} + \textbf{s}^T\nabla J(\boldsymbol{\theta}),
\label{eq: quadratic problem}
\end{equation}
where $B_{\textbf{D},\Delta}(\boldsymbol{\theta})=\{\textbf{s}:\left\Vert \textbf{D}(\textbf{s}-\boldsymbol{\theta})\right\Vert_2\leq\Delta\}$ is the trust region. The trust region is an ellipsoid with radius $\Delta$ and scaling matrix $\textbf{D}$ around the current parameter $\boldsymbol{\theta}$. The size of the trust region can be adapted over the course of iterations. Trust-region methods that do not use quadratic approximations use other local approximations, e.g., via radial basis functions~\cite{WildRom2008}.

Trust-region methods that solve the subproblem~\eqref{eq: quadratic problem} exactly are not attracted to saddle points and not limited to small step sizes~\cite{DauphinPas2014}. However, the quadratic problem \eqref{eq: quadratic problem} is usually difficult to solve exactly and is approximatively solved instead~\cite{NocedalWri2006}. For convex problems, the dogleg method~\cite{Steihaug1983}, which employs a linear combination of gradient and Newton step, can be applied. For non-convex problems, the two-dimensional subspace minimization method~\cite{ByrdSch1988,BranchCole1999} can be used. The two-dimensional subspace minimization method dampens the Hessian and can be seen as a trust-region variant of the Levenberg-Marquardt method. The dogleg and two-dimensional subspace minimization method both reduce the trust-region subproblem to a two dimensional problem, which renders the computational cost of determining the update step from a given gradient \textit{per se} independent of the number of parameters of the underlying problem. This feature makes them particularly suited for large-scale problems. However, the dampening of the Hessian can again lead to small step-sizes close to saddle points~\cite{DauphinPas2014}. 

\begin{table}[t]
\begin{center}
\caption{Implementations and interfaces of of optimization methods in popular computational biology toolboxes. Some toolboxes may feature variants of the cited algorithms. Some entries may be names of functions that feature multiple different algorithms. }
\label{tab: Implementation Optimization}
\begin{tabular}{|c||c|c|c|}
\hline
Toolbox	                   & Global                  & \multicolumn{2}{c|}{Local}    \\ 
\hline
         	               & 						 &  Derivative-Free & Gradient-Based   \\ 
\hline\hline
COPASI~\cite{HoopsSah2006} & Evolutionary Programming~\cite{FogelFog1991}& Nelder Mead~\cite{NelderMea1965}& Levenberg-Marquardt ~\cite{Levenberg1944,Marquardt1963}   \\ 
                           & Genetic Algorithm~\cite{Michalewicz2013} & Pattern Search~\cite{HookeJee1961} & Steepest Descent~\cite{NocedalWri2006}           \\ 
                           & Particle Swarm~\cite{Kennedy2011} &PRAXIS~\cite{Brent1973} & Truncated Newton~\cite{DemboSte1983}            \\
                           & Random Search~\cite{SolisWet1981}  & &           \\ 
                           & Simulated Annealing ~\cite{KirkpatrickGel1983}  & &           \\ 
                           & Scatter Search~\cite{Egea2007}  & &           \\
                           & SRES~\cite{RunarssonYao2000} & & \\
\hline
D2D~\cite{RaueSte2015}     & fminsearchbnd (MATLAB)  &                  & arNLS (custom)      \\ 
                           & Genetic Algorithm (MATLAB) &               & CERES (Google) \\ 
                           & multi-start (custom)    & 			        & fmincon (MATLAB)  \\
                           & Pattern Search (MATLAB) &                  & lsqnonlin (MATLAB)\\
                           & Particle Swarm (custom) &                  & STRSCNE~\cite{BellaviaMac2004}   \\
                           & Simulated Annealing (MATLAB)  &            & TRESNEI~\cite{MoriniPorc2012}  \\
\hline
MEIGO~\cite{EgeaHen2014}   & fminsearchbnd (MATLAB)  & DHC~\cite{DeLaMazaYur1994}   & fmincon (MATLAB) \\ 
                           & NOMAD~\cite{LeDigabel2011}& Pattern Search~\cite{kelley1999} & IpOpt~\cite{WachterBie2006}   \\ 
                           & multi-start(custom)   & SOLNP~\cite{Ye1989}& lsqnonlin (MATLAB)  \\ 
                           & Scatter Search~\cite{Egea2007} & 			& MISQP~\cite{ExlerLeh2012} \\ 
                           &                        &            		& N2FB~\cite{DennisGay1981} \\ 
                           &						&					& NL2SOL~\cite{DennisGay1981} \\
\hline

PESTO~\cite{StaporWei2018} &  multi-start (custom)   & BOBYQA~\cite{Powell2009} & fmincon (MATLAB) \\ 
                           &  Particle Swarm~\cite{VazVic2009}& DHC~\cite{DeLaMazaYur1994} & lsqnonlin (MATLAB)  \\ 
                           &  MEIGO~\cite{EgeaHen2014}& &   \\
\hline
\end{tabular}
\end{center}
\end{table}

\subsubsection{Implementation and Practical Considerations}

State-of-the-art parameter inference toolboxes for computational biology, such as D2D~\cite{RaueSte2015}, PESTO~\cite{StaporWei2018}, MEIGO~\cite{EgeaHen2014} and COPASI~\cite{HoopsSah2006}, feature a mix of local and global methods which include derivative-free and gradient-based methods (see Table~\ref{tab: Implementation Optimization}). In terms of global, derivative-free methods, most toolboxes provide interfaces to Particle Swarm and Pattern Search methods. In terms of local, gradient-based methods, most toolboxes feature various flavors of the trust-region algorithm. All MATLAB toolboxes provide interfaces to the fmincon and lsqnonlin routines from the MATLAB Optimization Toolbox. Only COPASI provides the implementation of more basic algorithms such as Levenberg-Marquardt~\cite{Levenberg1944,Marquardt1963}, Truncated Newton and Steepest Descent. In terms of global optimization schemes, all toolboxes employ either multi-start or scatter search algorithms~\cite{Egea2007}.

Choosing a particular optimization method from this plethora of choices is not an easy task. There are no exhaustive studies that compare the full range of different optimization methods on a large set of problems. In some studies, gradient-based optimization algorithms perform best~\cite{RaueSch2013,HrossHas2016}, but others also show that derivative-free methods can perform well~\cite{DegasperiFey2017}. In general, a rigorous evaluation of optimization algorithms is highly involved, as there are small differences in the implementations of various algorithms. For example, STRSCNE~\cite{BellaviaMac2004}, RESNEI~\cite{MoriniPorc2012}, NL2SOL~\cite{DennisGay1981}, lsqnonlin and fmincon all implement trust-region algorithms and even for expert users it may be difficult to pinpoint differences between individual implementations. A recent study suggests substantial differences in the efficiency of various implementations of trust-region algorithms and identified lsqnonlin to be the best performing algorithm~\cite{Wieland2016}. Even for a single implementation the specification of hyper-parameters can have substantial impact on performance. Many algorithms require user specifications of technical parameters. Finding good values for these hyper-parameters may be challenging for non-expert users and default values may not work for all problems. 

As few researchers are experts in a large number of different optimization methods, a rigorous evaluation of multiple different algorithms is challenging. To circumvent this problem, a recent study~\cite{VillaverdeHen2015} suggested a set of benchmark problems on which other researchers are invited to evaluate their algorithms. Yet, few algorithms have been evaluated on that benchmark so far~\cite{PenasGon2015,FroehlichKal2017}. Complementarily, \citep{Kreutz2016} suggests the construction of statistical models to assess the performance of methods and the effect of hyper-parameters.

The different optimization algorithms we outlined in this section rely on evaluations of the objective function, its gradient or even its Hessian. In the following sections we will the discuss methods to evaluate these terms.

\subsection{Simulation}

The objective function and its gradient are typically not available in closed form, but have to be computed numerically. For large-scale models the computational cost of computing the objective function and its gradient is high, which makes parameter estimation computationally demanding. Depending on the class of the employed model and simulation algorithm, the computation time will depend on different features of the underlying model, which we will discuss in detail in the following.

The timescales of biochemical processes span multiple orders of magnitude~\cite{ShamirBar2016,HasenauerJag2015}. As comprehensive models often cover a large variety of different biological processes, they are particularly prone to possess multiple timescales~\cite{SmallboneMen2013}. This results in the stiffness of corresponding ODEs~\cite{ResatPet2009}. As the stiffness of the equations typically depends on the choice of parameters, it is rarely possible to assess the stiffness \textit{a priori}. Consequently, it is always advisable to use implicit solvers, which can adequately handle stiffness, for parameter inference~\cite{GonnetDim2012}. 

\subsubsection{Implicit Methods}

For stiff problems, implicit differential equation solvers from the fully implicit Runge-Kutta solver family~\cite{Butcher1964}, the Singly Diagonally Implicit Runge-Kutta solver family~\cite{Alexander1977} or the Rosenbrock solver family~\cite{Rosenbrock1963} should be used. These solvers compute the state variables at the next time step $\xi_{i}$ based on the state variables at previous iterations $x(\xi_{i-1}),x(\xi_{i-2}),x(\xi_{i-3}),\ldots$ by solving an implicit equation
\begin{equation*}
G(\textbf{x}(\xi_{i}),\textbf{x}(\xi_{i-1}),\textbf{x}(\xi_{i-2}),\textbf{x}(\xi_{i-3}),\ldots)=0,
\end{equation*}
where the function $G$ depends on the choice of the method and on the right hand side of the differential equation $f$.

For single-step methods, such as the Runge-Kutta type solvers, the function $G$ will only depend on $x(\xi_{i-1})$, and not on previous values. For implicit Runge-Kutta solvers, a system of linear equations with $\dims{x}\cdot s$ equations has to be solved in every iteration~\cite{ZhangAdr2014}. Here  $s$ is a the number of stages, which is a particular property of a Runge-Kutta solver which determines the order of the method.

For multi-step methods, the function $G$ will also depend on previous values of $\textbf{x}$. A popular implementation of the multi-step method is the implicit linear multi-step Backwards Differentiation Formula (BDF) implemented in the CVODES solver~\cite{SerbanHin2005}. In every iteration $i$, the BDF solves an equation of the form
\begin{equation*}
h_i\beta_{i,0}f(\xi_{i},\textbf{x}(\xi_{i}),\boldsymbol{\theta})+\sum_{j=0}^q \alpha_{i,j}\textbf{x}(\xi_{i-j}) =0,
\end{equation*}
where $q$ is the order of the method, $\boldsymbol{\alpha}$ and $\boldsymbol{\beta}$ are the coefficients that are determined in every iteration. The order $q$ and the step size $h_i$ will determine the local error of the numerical solution~\cite{SerbanHin2005} and are often chosen adaptively. This implicit equation is typically solved using Newton's method~\cite{HindmarshBro2005,ZhangAdr2014}. For the BDF, the function $G$ depends on $\textbf{$\dot{x}$}$  and thus on $f$. Consequently, the Newton solver computes multiple solutions to linear systems defined by the Jacobian $\nabla_x f(t,\textbf{x},\boldsymbol{\theta})$ of the right hand side of the differential equation at every integration step. As this linear system only has $\dims{x}$ equations, in contrast to the $\dims{x}\cdot s$ equations for single-step methods, the computational cost of solving the linear system increases less strongly with the number of state variables $\dims{x}$.

The computation time of the BDF method primarily depends on two factors: (i) the evaluation time of the function $f$ and the Jacobian $\nabla_x f(t,\textbf{x},\boldsymbol{\theta})$, which usually scales linearly with $\dims{x}$ and (ii) the time to solve the linear systems defined by $\nabla_x f(t,\textbf{x},\boldsymbol{\theta})$. The matrix $\nabla_x f(t,\textbf{x},\boldsymbol{\theta})$ is typically not symmetric and neither positive nor negative definite. For such unstructured problems, LU decomposition (see Figure~\ref{fig:optimization}C middle), which factorizes $\nabla_x f(t,\textbf{x},\boldsymbol{\theta})$ into a lower-triangular matrix $\textbf{L}$ and an upper-triangular matrix $\textbf{U}$, is the method of choice to solve the linear system, as long as $\textbf{L}$ and $\textbf{U}$ can be stored in memory. After performing the decomposition, the solution to the linear systems can be computed by matrix multiplication. When no additional structure of the matrix is exploited, the computational complexity of matrix multiplication with state-of-the-art algorithms increases at least with exponent $2.376$ with respect to $\dims{x}$~\cite{CoppersmithShm1990} and thus dominates the computation time for sufficiently large $\dims{x}$.

\subsubsection{Sparse Implicit Methods}

For ODE models arising from discretization of partial differential equations, the Jacobian can usually be brought into banded form. For such banded matrices, specialized solvers that scale with the number of off-diagonals of the Jacobian have been developed~\cite{Thorson1979}. Unfortunately, ODE models of biochemical reaction networks cannot generally be brought into a banded structure. For example, in polymerization reactions that include dissociation of monomers, the monomer species will always be influenced by all other species and the number of off-diagonals in the Jacobian will be equal to $\dims{x}$. Other frequently occurring motifs, such as feedback loops and single highly interactive species~\cite{BarabasiOlt2004}, will also increase the number of necessary off-diagonals. 

As alternative to banded solvers, sparse solvers have been introduced in the context of circuit simulations~\cite{DavisPal2010}. For sparse solvers, the computation time depends on the number of non-zero entries in $\nabla_x f(t,x,\boldsymbol{\theta})$\makered{, which scales with the number of biochemical reactions}. The sparse solver relies on an approximate minimum degree (AMD) ordering~\cite{AmestoyDav1996} which is a graph theoretical approach that can be used to minimize the fill-in of the \textbf{L} and \textbf{U} matrices of the LU-decomposition (see ~\ref{fig:optimization}C right). Currently, no formulas for the expected speedup or the general scaling with respect to non-zero entries exist. For biochemical reaction networks, the application of such a sparse solver seems reasonable~\cite{GonnetDim2012}, but no rigorous evaluation of the scaling has been performed.

\subsubsection{Implementation and Practical Considerations}

Most toolboxes use CVODES~\cite{SerbanHin2005} or LSODA~\cite{Petzold1983} for simulation (see Table~\ref{tab: Implementation Simulation}). In contrast to CVODES, which only implements an implicit solver, LSODA dynamically switches between explicit and implicit solvers. To the best of our knowledge, no comparison between LSODA and CVODES has ever been published. However, LSODA does not provide an interface to the Clark Kent LU solver (KLU)~\cite{DavisPal2010} or any other sparse solver and thus might perform poorly on large-scale problems with sparsity structure. Another notable difference is that only few toolboxes analytically compute the Jacobian of the right hand side and provide it to the solver. The symbolic processing is necessary for the sparse representation, but also likely to be beneficial for dense solvers. Thus, D2D~\cite{RaueSte2015} and AMICI~\cite{FroehlichKal2017} are also the only general purpose simulation libraries for systems biology that allow the use of the sparse KLU solver.

For explicit solvers, no linear system has to be solved and the algorithm largely consists of elementary operations which can be efficiently parallelized on GPUs~\cite{TangherloniNob2017}. For explicit solvers, also parallelized solvers are available~\cite{DemmelGil1999}. However, the computational overhead of parallelization is usually too high, unless models with several thousand state variables are considered~\cite{DemmelGil1999}.

All of the considered toolboxes allow the definition of ODE models in the Systems Biology Markup Language (SBML)~\cite{HuckaFin2003}. This also allows for the definition of models in terms of biochemical reactions. Here COPASI and libRoadRunner aim for a full support of SBML features, while AMICI and D2D only support a subset of SBML features. 

\begin{table}[t]
\begin{center}
\caption{Implementations of simulation methods in popular computational biology toolboxes.}
\label{tab: Implementation Simulation}
\begin{tabular}{|c||c|c|c|c|}
\hline
Toolbox	                   & Simulation Library & Jacobian    & Dense Solver                  & Sparse Solver    \\ 
\hline\hline
AMICI~\cite{FroehlichKal2017}& CVODES~\cite{SerbanHin2005} & symbolic & \checkmark       & \checkmark     \\  
\hline
COPASI~\cite{HoopsSah2006} & LSODA~\cite{Petzold1983} & numeric  & \checkmark       & $\times$     \\  
\hline
D2D~\cite{RaueSte2015}     & CVODES~\cite{SerbanHin2005} & symbolic & \checkmark       & \checkmark     \\ 
\hline
libRoadRunner~\cite{SomogyiBou2015} & CVODES~\cite{SerbanHin2005} & numeric  & \checkmark       & $\times$     \\  
\hline

\end{tabular}
\end{center}
\end{table}

Sparse numerical solvers can be used to efficiently compute the numerical solution to the ODE, which are required for objective function evaluation. They can also be used to compute objective function gradients as solution to one or more ODEs. Several different gradient computation approaches exist and in the following we will discuss the three most common approaches.

\subsection{Gradient Calculation}

Providing an accurate gradient to the objective function is essential for gradient-based methods~\cite{RaueSch2013,GriewankWal2008,NocedalWri2006}. For ODE constrained optimization problems, the gradient of the objective function can be computed based on the parametric derivative of the solution to the ODE. These derivatives are often called the sensitivities of the model. Several approaches to compute sensitivities for ODE models exist, including finite differences~\cite{Milne1933}, as well as the direct approach via forward sensitivity analysis~\cite{DickinsonGel1976,KokotovicHel1967}.

\subsubsection{Finite Differences and Forward Sensitivity Analysis}

For finite differences, the entries of the gradient are approximated according to 
\begin{equation*}
\frac{\mathrm{d} J}{\mathrm{d} \theta_k} \approx \frac{J(\boldsymbol{\theta}+a \, \textbf{e}_k) - J(\boldsymbol{\theta}-b \, \textbf{e}_k)}{a+b},
\end{equation*}
with $a,b \geq 0$ and the $k^{th}$ unit vector $\textbf{e}_k$. In practice, forward differences ($a = \zeta$, $b = 0$), backward differences ($a = 0$, $b = \zeta$) and central differences ($a = \zeta$, $b = \epsilon$), with $\zeta \ll 1$, are widely used. As the evaluation of $J(\boldsymbol{\theta}+a \, \textbf{e}_k)$ and $J(\boldsymbol{\theta}-b \, \textbf{e}_k)$ may require additional solutions to the model ODE, the scaling of finite differences with respect to the number of parameters is also linear.

For forward sensitivity analysis, the entries of the gradient of the objective function are computed according to
\begin{align*}
\frac{\mathrm{d} J}{\mathrm{d} \theta_k}  =  -\sum_{i=1}^\dims{y} \sum_{j = 1}^{T} \frac{\partial J}{\partial y_i(t_j,\theta)} s_{i,k}^y(t_j,\boldsymbol{\theta}) +  \frac{\partial J}{\partial \theta_k},
\end{align*}
with $s_{i,k}^y(t,\boldsymbol{\theta})$ denoting the sensitivity of output $y_i$ at time point $t_j$ with respect to parameter $\theta_k$. This output sensitivity can be computed by applying the total derivative to the functions $h$:
\begin{align*}
s^y_{i,k}(t_j,\boldsymbol{\theta}) = \left.\frac{\partial h_i}{\partial x}\right|_{\textbf{x}(t,\boldsymbol{\theta}),\boldsymbol{\theta}}  s^x_k(t_j,\boldsymbol{\theta}) + \left.\frac{\partial h_i}{\partial \theta_k}\right|_{x(t,\boldsymbol{\theta}),\boldsymbol{\theta}}
\end{align*}
with $\textbf{s}^x_k(t,\boldsymbol{\theta})$ denoting the sensitivity of the state $\textbf{x}$ with respect to $\theta_k$. The state sensitivity is defined as solution to the ODE system:
\begin{equation*}
\dot{\textbf{s}}^x_k(t,\boldsymbol{\theta}) =  \left.\frac{\partial f}{\partial x}\right|_{x(t,\boldsymbol{\theta}),\boldsymbol{\theta}} \textbf{s}^x_k(t,\boldsymbol{\theta}) + \left.\frac{\partial f}{\partial \theta_k}\right|_{x(t,\boldsymbol{\theta}),\boldsymbol{\theta}} , \qquad \textbf{s}^x_k(t_0,\boldsymbol{\theta}) = \left.\frac{\partial x_0}{\partial \theta_k}\right|_{\boldsymbol{\theta}}. \label{eq: dynamical system - fse} 
\end{equation*}
Thus, forward sensitivity analysis requires the computation of a solution to an ODE system of the same size as the model ODE for every gradient entry. Consequently, the scaling with respect to the number of parameters is linear (see Figure~\ref{fig:optimization}B right).

\subsubsection{Adjoint Sensitivity Analysis}

The linear scaling of forward sensitivity analysis and finite differences can be computationally prohibitively demanding for large-scale models with thousands of parameters. The alternative adjoint approach, which computes the objective function gradient via adjoint sensitivity analysis, has long been deemed to be computationally more efficient for systems with many parameters~\cite{KokotovicHel1967}. In other research fields, e.g., for partial differential equation constrained optimization problems, adjoint sensitivity analysis~\cite{HindmarshBro2005} has been adopted in the past decades. In contrast, in the systems biology community there are only isolated applications of adjoint sensitivity analysis~\cite{LuMuller2008,FujarewiczKimmel2005,LuAug2012}.

In the mathematics and engineering community, adjoint sensitivity analysis is frequently used to compute the gradients of a functional with respect to the parameters if the functional depends on the solution of a differential equation~\cite{Plessix2006}. In these applications, measurements are continuous in time and $J(\boldsymbol{\theta})$ is assumed to be a functional of the solution $\textbf{x}(t)$ of a differential equation. However, this approach can also be applied to discrete-time measurements and in contrast to forward sensitivity analysis, adjoint sensitivity analysis does not rely on the state sensitivities $\textbf{s}^x_k(t)$, but on the adjoint state $\textbf{p}(t)$.

For discrete-time measurements -- the usual case in systems and computational biology -- the adjoint state is piece-wise continuous in time and defined by a sequence of backward differential equations~\cite{FroehlichKal2017}. For $t > t_N$, the adjoint state is zero, $\textbf{p}(t) = 0$. Starting from this end value, the trajectory of the adjoint state is calculated backwards in time, from the last measurement $t = t_N$ to the initial time $t = t_0$. At the measurement time points $t_N, \ldots, t_1$, the adjoint state is reinitialized as
\begin{equation}
\textbf{p}(t_j) = \lim_{t \rightarrow t_j^+} \textbf{p}(t) + \frac{\partial J}{\partial x},
\label{eq: end value for intervals}
\end{equation}
which usually results in a discontinuity of $p(t)$ at $t_j$. Starting from the end value $p(t_j)$ as defined in~\eqref{eq: end value for intervals} the adjoint state evolves backwards in time until the next measurement point $t_{j-1}$ or the initial time $t_0$ is reached. This evolution is governed by the time dependent linear ODE
\begin{equation}
\dot{\textbf{p}} = - \left(\frac{\partial f}{\partial x}\right)^T \textbf{p}.
\label{eq: end value problem for intervals}
\end{equation}
The repeated evaluation of~\eqref{eq: end value for intervals} and~\eqref{eq: end value problem for intervals} until $t = t_0$ yields the trajectory of the adjoint state. Given this trajectory, the gradient of the objective function with respect to the individual parameters is 
\begin{equation}
\frac{\mathrm{d} J}{\mathrm{d} \theta_k} = - \int_{t_0}^{t_N} \textbf{p}^T \frac{\partial f}{\partial \theta_k} dt - \textbf{p}(t_0)^T \frac{\partial \textbf{x}_0}{\partial \theta_k} + \frac{\partial J}{\partial \theta_k}.
\label{eq: integral for gradient calculations using the adjoint}
\end{equation}
The key advantage of this approach is that \eqref{eq: integral for gradient calculations using the adjoint}, which has to be evaluated for every parameter separately, can be evaluated very efficiently, while \eqref{eq: end value problem for intervals}, which is computationally more demanding, only has to be solved once~\cite{FroehlichKal2017}. As \eqref{eq: end value problem for intervals} is of the same dimensionality as the original ODE model, this allows the computation of gradients at the cost of roughly two solutions of the original ODE model. In practice the adjoint sensitivity approach has an almost constant scaling with respect to the number of parameters.

\subsubsection{Implementation and Practical Considerations}

Many toolboxes rely on finite differences to compute gradients (see Table~\ref{tab: Implementation Gradient}). D2D~\cite{RaueSte2015} and AMICI~\cite{FroehlichKal2017} are two notable examples that allow the computation of gradients via sensitivity analysis, but only AMICI allows adjoint sensitivity analysis.

\begin{table}[t]
\begin{center}
\caption{Implementations of gradient computation methods in popular computational biology toolboxes.}
\label{tab: Implementation Gradient}
\begin{tabular}{|c||c|c|c|c|}
\hline
Toolbox	                   & Finite Differences & Forward Sensitivity    & Adjoint Sensitivity    \\ 
\hline\hline
AMICI~\cite{FroehlichKal2017}& $\times$ & \checkmark & \checkmark            \\  
\hline
COPASI~\cite{HoopsSah2006} & \checkmark & $\times$  & $\times$           \\  
\hline
D2D~\cite{RaueSte2015}     & \checkmark & \checkmark & $\times$           \\ 
\hline
libRoadRunner~\cite{SomogyiBou2015} & \checkmark & $\times$  & $\times$          \\  
\hline

\end{tabular}
\end{center}
\end{table}

\subsection{Hessian Computation}

In addition to the gradient, Newton-type methods also require the Hessian $\nabla^2J(\boldsymbol{\theta})$ of the objective function. The numerical evaluation of the Hessian can be challenging as the dependence of the computational complexity on the number of parameters $\dims{\theta}$ is one order higher than for the gradient: The computation time for finite differences and forward sensitivities scale quadratically with the number of parameters~\cite{BalsaBan2001,VassiliadisCan1999}. For adjoint sensitivities, the computation time depends linearly on the number of parameters~\cite{OzyurtBar2005}.

\subsubsection{Gauss-Newton Approximation}

For independent, normally distributed measurement noise, as assumed in~\eqref{eq: objective function}, and known noise parameters $\sigma$, the optimization problem~\eqref{eq: optimization problem} is of least squares type. This structure can be exploited by using Gauss-Newton (GN)~\cite{bjorck1996} type algorithms, which ignore second order partial derivatives in the Hessian. The respective approximations of the Hessian coincide with the Fisher information matrix (FIM)~\cite{Fisher1922} of the respective parameter estimate. The key advantage of this approach is that the FIM can be computed for the same cost as one gradient using forward sensitivity analysis.

\subsubsection{Quasi-Newton Approximation}

For problems that are not of least-squares type, quasi-Newton methods such as the Broyden-Fletcher-Goldfarb-Shanno (BFGS)~\cite{FletcherPow1963,Goldfarb1970} algorithm can be used. The BFGS algorithm iteratively computes approximations to the Hessian based on updates which are derived from the outer products of the gradient $\nabla J(\boldsymbol{\theta})$ of the objective function. The resulting approximation is guaranteed to be positive definite, as long as the Wolfe condition~\cite{Wolfe1969} is satisfied in every iteration and the initial approximation is positive definite. As previously discussed, positive definiteness ensures descent directions for line-search methods, but will generally lead to small step sizes in the vicinity of saddle points. The symmetric rank 1 (SR1) algorithm addresses this problem by allowing for negative- and indefinite approximations~\cite{ByrdKha1996}. This procedure facilitates the application of optimization methods which avoid saddle points by allowing directions of negative curvature~\cite{RamamurthyDuf2017}. 

\makered{Quasi-Newton versions are generally cheap to compute, as they only require simple algebraic manipulations of the gradient. Algorithms based on limited memory variants such as L-BFGS~\cite{Nocedal1980,LiuNoc1989} or L-SR1~\cite{RamamurthyDuf2017} have been applied to machine learning problems with millions of parameters~\cite{AndrewGao2007}.}

\subsubsection{Implementation and Practical Considerations}

\begin{table}[t]
\caption{Implementations of (approximative) Hessian computation methods in popular computational biology toolboxes. For BFGS and SR1 we list the function or option that allows the respective approximation. SE = sensitivity equations, FD = finite differences, GN = Gauss-Newton, N/A = Not applicable}
\label{tab: Implementation Hessian}
\begin{center}
\begin{tabular}{|c||c|c|c|c|}
\hline
Toolbox	                   & FIM / GN & Hessian & BFGS & SR1 \\ 
\hline\hline
AMICI~\cite{FroehlichKal2017} & SE & SE  & N/A & N/A \\
\hline
COPASI~\cite{HoopsSah2006} & FD & $\times$ & Truncated Newton & $\times$ \\  
\hline
D2D~\cite{RaueSte2015}     &  SE/FD & $\times$ & fmincon & arNLS\_SR1 \\ 
\hline
libRoadRunner~\cite{SomogyiBou2015} & $\times$  & $\times$ & N/A & N/A \\  
\hline
MEIGO~\cite{HoopsSah2006} & N/A & N/A & fmincon, IpOpt & IpOpt \\  
\hline
PESTO~\cite{StaporWei2018}&  N/A & N/A  & fmincon & $\times$ \\ 
\hline
\end{tabular}
\end{center}
\end{table}

\makered{The implementation of methods for the computation of (approximate) Hessians is quite disparate across toolboxes (see Table~\ref{tab: Implementation Hessian})}. AMICI is the only toolbox that allows sensitivity-based computation of the Hessian. Most other toolboxes use FIM/Gauss-Newton, BFGS or SR1 approximations. Most of iterative approximations BFGS and SR1 are implemented as part of the optimization algorithm and it is not possible to use them with other methods. Only D2D provides a relatively flexible implementation of SR1. For the FIM approximation and the exact Hessian computations, the implementations are usually transferable between optimization methods. 
\makered{In theory, the computation of the exact Hessian, with adjoint sensitivity analysis, and the FIM, with forward sensitivity analysis, both scale linearly with the number of parameters and only the exact Hessian can be used to construct methods that avoid saddle points~\cite{DauphinPas2014}. In practice, the effect of using the FIM over the Hessian on the efficiency of respective optimization methods has not been studied for systems biology problems. For problems with thousands of parameters, the computation of both may be challenging and the BFGS and SR1 approximations become more appealing.}

In this section, we split the parameter inference problem in three parts: optimization, simulation and gradient computation and discussed respective scaling properties. These techniques generalize to other model analysis techniques that require optimization or gradient computation, such as uncertainty analysis~\cite{Raue2009,GirolamiCal2011}, experimental design~\cite{VanlierTie2012} and the inference of model structure. A detailed discussion of all these methods is beyond the scope of this book chapter, but in the following we will discuss the inference of model structure in more detail.

\section{Inference of Model Structure}
In many applications, it is not apparent which biochemical species and reactions are necessary to describe the dynamics of a biochemical process. In this case, the structure of the ODE model~\eqref{eq: ODE}, i.e., vector field $f(\textbf{x},\boldsymbol{\theta})$ and initial condition $\textbf{x}_0(\boldsymbol{\theta})$, have to be inferred from experimental data. The selection should compromise between goodness-of-fit and complexity. Following the concept of Occam's razor~\cite{BlumerEhr1987}, one tries to control variability associated with over-fitting while protecting against the bias associated with under-fitting. 

In the following, we formulate the problem of model structure inference. We introduce and discuss criteria that select models out of a set of candidate models and describe approaches to reduce the number of candidate models. We outline the scalability of the approaches and their computational complexity.

\subsection{Model Selection Criteria}
\label{sec:model_selection}
Given a set of candidate models $M_1,M_2,\ldots,M_{\dims M}$ the aim of model inference is to find a model or a set of models which (i) describe the data available and (ii) generalize to other datasets~\cite{Hastie2009}. The choice of model can be made with several selection criteria, differing among others in asymptotic consistency~\cite{Shibata1980}, asymptotic efficiency~\cite{Fisher1922} and computational complexity. If the true model is included in the set of candidate models, a consistent criterion will asymptotically select the true model with probability one and an efficient criterion will select the model that minimizes the mean squared error of the prediction. 

While the concepts in the previous sections followed the frequentist approach, some of the concepts presented in this section are Bayesian. For these approaches, prior knowledge about the parameters is incorporated and the posterior probability~\eqref{eq: poster} is analyzed instead of the likelihood function. 

One popular criterion is the Bayes factor~\cite{KassRaf1995}, which has been shown to be asymptotically consistent for a broad range of models (e.g.,~\cite{WangSun2014,ChoiRou2015}), however, for the case of general ODE models, no proofs for asymptotic efficiency and consistency are available for all the criteria presented in this section. Bayes' Theorem yields the posterior model probability 
\begin{align}
p(M_m|\mathcal{D}) = \frac{p(\mathcal{D}|M_m)p(M_m)}{p(\mathcal{D})}
\end{align}
with marginal likelihood 
\begin{align}
p(\mathcal{D}|M_m) = \int_{\Theta_m} p(\mathcal{D}|\boldsymbol{\theta}_m)p(\theta_m|M_m)d\boldsymbol{\theta}_m
\end{align}
with model prior $p(M_m)$ and marginal probability $p(\mathcal{D}) = \sum_j p(\mathcal{D}|M_j) P(M_j)$.
The Bayes factor of models $M_1$ and $M_2$ is the ratio of the corresponding marginal likelihoods
\begin{equation}
B_{12} = \frac{p(\mathcal{D}|{M}_1)}{p(\mathcal{D}|{M}_2)}.
\end{equation}
The Bayes factor describes how much more likely it is that the data are generated from $M_1$ instead of $M_2$. A Bayes factor $B_{12}>100$ is often considered decisive for rejecting model ${M}_2$~\cite{Jeffreys1961}. The Bayes factor intrinsically penalizes model complexity by integrating over the whole parameter space of each model. Bayes factors can be approximated by Laplace approximation, which has a low computational complexity but provides only a local approximation. To enable a more precise computation of the Bayes factors, bridge sampling~\cite{MengWon1996}, nested sampling~\cite{Skilling2006}, thermodynamic integration~\cite{HugSch2016}, or related methods can be employed. These approaches evaluate the integral defining the marginal likelihood $p(\mathcal{D}|M_m)$. As the approaches require a large-number of function evaluations, the methods are usually computationally demanding and the computational complexity is highly problem-dependent. Thus, efficient sampling methods are required.

For high-dimensional or computationally demanding problems, the calculation of Bayes factors might be intractable and computationally less expensive model selection criteria need to be employed. A model selection criterion which is based on the MLE, instead of a marginal likelihood (an integral over the whole parameter space), is the Bayesian Information Criterion (BIC)~\cite{Schwarz1978}. The BIC value for model $M_m$ is
\begin{align}
\mathrm{BIC}_m &= -2\log\left(p\left(\mathcal{D}|\boldsymbol{\theta}_m^*\right) \right) + \log\left(|\mathcal{D}|\right)\dims{\theta_m}\,.\label{eq:BIC}
\end{align}
For structural identifiable models, the BIC provides in the limit of large sample sizes information about the Bayes factors,
\begin{align}
\lim_{|\mathcal{D}|->\infty} \frac{-2\log B_{12} - (\mathrm{BIC}_1 - \mathrm{BIC}_2)}{-2\log B_{12}} = 0\,.
\end{align}
From information theoretical arguments, the Akaike Information Criterion (AIC)
\begin{align}
\mathrm{AIC}_m &= -2\log\left(p\left(\mathcal{D}|\boldsymbol{\theta}_m^*\right) \right) + 2\dims{\theta_m}\,,\label{eq:AIC}
\end{align}
has been derived~\cite{Akaike1973}. Low BIC and AIC values are preferable and differences above $10$ are assumed to be substantial (see Table~\ref{tab:thresholds} and \cite{KassRaf1995,BurnhamAnd2002}).

For model selection in many problem classes, the AIC is asymptotically efficient, but not consistent, while the BIC is asymptotically consistent, but not efficient~\cite{Shibata1981,Kuha2004,Acquah2010}.

\begin{table}[t]
\begin{center}
\caption{Decisions based on the Bayes factor and differences in BIC and AIC values~\cite{Jeffreys1961,KassRaf1995,BurnhamAnd2002}.}
\label{tab:thresholds}
\begin{tabular}{|c|c|c||c|}
\hline
B$_{lm}$ & $\mathrm{BIC}_m -\min_l \mathrm{BIC}_l$ & $\mathrm{AIC}_m -\min_l \mathrm{AIC}_l$  & decision  \\ \hline\hline
$1-3$   &  $0-2$ & $0-4$  & do not reject model $M_m$  \\ \hline
$3-100$ &  $2-10$ & $4-10$  & -  \\ \hline
$>100$ &  $>10$ &  $>10$ & reject model $M_m$  \\ \hline
\end{tabular}
\end{center}
\end{table}

When incorporating prior information about parameters, the priors can conceptually be treated as additional data points and, thus, be part of the likelihood to still allow the use of BIC and AIC. Also extensions of the criteria exist, such as the corrected AIC~\cite{HurvichTsi1989}, which provides a correction for finite sample sizes. Also other extended versions of the criteria have been developed (see, e.g., \cite{ChenChe2008}), however, the discussion of these is beyond the scope of this chapter.

For the comparison of nested models $M_m$ and $M_l$ , i.e., $\boldsymbol{\theta}_m \in \Theta_m$ and $\boldsymbol{\theta}_l \in \Theta_l$ where $\Theta_m$ is a subset of $\Theta_l$, the likelihood ratio test can be applied \cite{Wilks1938}, which is an efficient test~\cite{NeymanPea1992}. The likelihood ratio is defined as
\begin{align}
\Lambda = \frac{p(\mathcal{D}|\boldsymbol{\theta}_m^*)}{p(\mathcal{D}|\boldsymbol{\theta}_l^*)} \leq 1\,,
 \end{align}
and model $M_m$ is rejected if $\Lambda$ is below a certain threshold which is obtained using Wilks' Theorem~\cite{Wilks1938}. This theorem states that it holds
in the large sample limit (see~\cite{Wilks1938} for further details)
\begin{align}
2(\log p(\mathcal{D}|\boldsymbol{\theta}_l^*)-\log p(\mathcal{D}|\boldsymbol{\theta}_m^*)) \sim \chi^2(\cdot|\dims{\theta_l}- \dims{\theta_m}).
\end{align}
Given a certain $\alpha$ level, model $M_m$ is rejected if
\begin{align}
\int_{0}^{2(\log p(\mathcal{D}|\boldsymbol{\theta}_l^*)-\log p(\mathcal{D}|\boldsymbol{\theta}_m^*))} \chi^2(\psi|\dims{\theta_l}- \dims{\theta_m})d\psi \geq 1-\alpha\,.
\end{align}While only Bayes factors and the likelihood ratio test are proven to be valid for non-identifiable parameters, the use of AIC and BIC can be problematic for these cases. 

The discussion of further criteria, such as the log-pointwise predictive density~\cite{GelmanHwa2014} or cross-validation (see, e.g.,~\cite{ArlotCel2010}), which evaluate the predictive quality of the model, is beyond the scope of this chapter. 

\subsubsection{Implementation and Practical Considerations}

AIC and BIC are rather simple to compute and are, among others, available in PESTO~\cite{StaporWei2018}. From the previously discussed toolboxes of this review, PESTO also provides sampling methods that can be employed to calculate Bayes factors, such as parallel tempering. Other toolboxes which can be employed for computing Bayes factors are, amongst others, BioBayes~\cite{VyshemirskyGir2008}, MultiNest~\cite{FerozHob2009}, or the C++ toolbox BCM~\cite{ThijssenDij2016}.

\subsection{Reduction of Number of Models}
\begin{figure}[tb]
\centering
\includegraphics[scale=0.4]{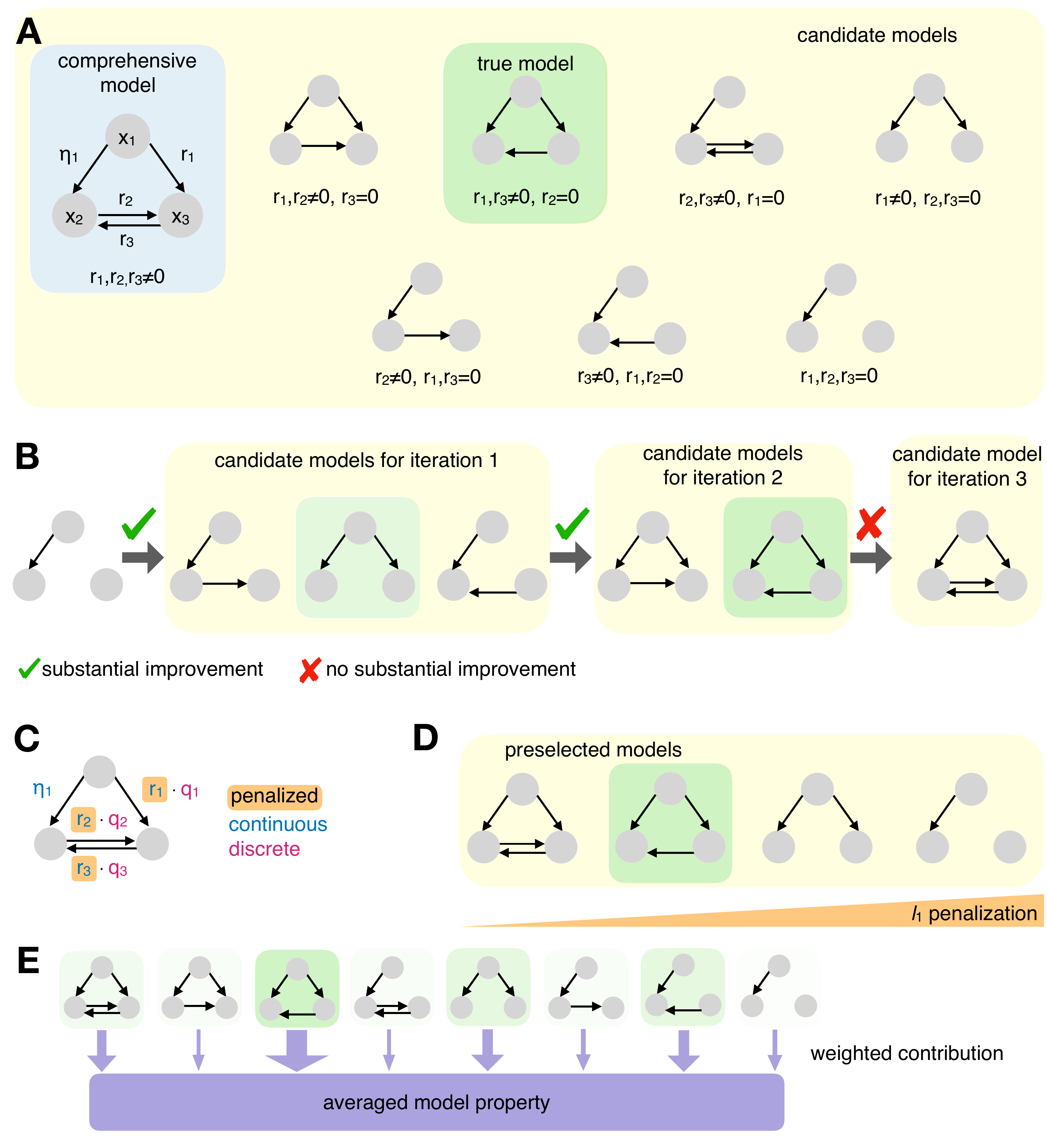}
\caption{Illustration of methods for model reduction. (A)~Set of candidate models, varying in the existence of connections between nodes $x_1, x_2$ and $x_3$. In total, there are $2^{\dims r} = 2^3$ models with at least $\dims{\eta} = 1$ parameters. (B)~Illustration of forward-selection starting from minimal model. In the first iteration, the model with $r_1\neq 0,r_2,r_3=0$ is selected (green) and in the second iteration the model with $r_1,r_3\neq 0,r_2=0$. The full model is rejected based on the selection criteria. (C)~To apply \textit{$l_0$ penalization} a MINLP problem needs to be solved, comprising continuous parameters $\boldsymbol{\eta}$ and $\mathbf{r}$ and discrete parameters $\textbf{q}\in \{0,1\}^{\dims r}$. (D)~\textit{$l_1$ penalization} reduces the number of potential models to a set of preselected models by increasing the penalization and thus forcing parameters $r$ to zero. (E)~Illustration of model averaging. The thickness of the arrows corresponds to the posterior probability, Akaike weight, or BIC weight and indicates the contribution of the model to the averaged model properties.}\label{fig:modelreduction}
\end{figure}
For most models, computing Bayes' factors is computationally demanding compared to optimization and the evaluation of AIC, BIC, or likelihood ratio. Yet, if the number of candidate models $\dims M$ is large, even the evaluation of AIC and BIC can become limiting as $\dims M$ optimization problems have to be solved. For non-nested models, the model selection criterion of choice needs to be calculated for each model to determine the optimal model.  

In this section, we consider a nested set of candidate models. In this case, all candidate models are a special case of a comprehensive model and can be constructed by fixing a subset of the parameters to specific values (Figure~\ref{fig:modelreduction}A). For the remainder of this chapter, we will assume that we can split the model parameters $\boldsymbol{\theta}$ into general parameters $\boldsymbol{\eta} \in \mathbb{R}^{\dims{\eta}}$, which are present in all models, and difference parameters $\mathbf{r} \in \mathbb{R}^{\dims r}$, which encode the nesting between models. Moreover, without loss of generality, it is assumed that $r_i = 0, i=1,\ldots,\dims{r}$ corresponds to the simplest model and $r_i \neq 0, i=1,\ldots,\dims{r}$ corresponds to the most complex model (see Figure~\ref{fig:modelreduction}A). These difference parameters could for example be the kinetic rates of hypothesized reactions~\cite{KlimovskaiaGan2016} or scaling factors for possibly cell-type or condition-specific parameters (see, e.g.,~\cite{SteiertTim2016}). Such settings yield a total of $2^{\dims{r}}$ candidate models, where $\dims{r}$ is limited by $\dims{\theta}$. Thus, for models with a high number of parameters, also a high number of nested models is possible. When $\dims{r}$ and $\dims{\theta}$ are both high, the inference of model parameters and thus the inference of model structure is challenging. 

\subsubsection{Forward-Selection and Backward-Elimination}
In statistics, step-wise regression is an often-used approach to reduce the number of models that need to be tested. This comprises forward-selection and backward-elimination (e.g.,~\cite{Hastie2009}) and combinations of both~\cite{KaltenbacherOff2011}. Forward-selection is a bottom-up approach which starts with the least complex model and successively activates individual difference parameters (i.e., setting $r_i\neq 0$) until a sufficient agreement with experimental data is achieved, evaluated using a model selection criterion (Figure~\ref{fig:modelreduction}B). In contrast, backward-elimination is a top-down approach starting with the most complex model, successively deactivating individual difference parameters (i.e., setting $r_i=0$) that are not required for a good fit to the data.

Forward-selection and backward-elimination reduce the number of models that need to be compared with the model selection criteria described before from $2^{\dims r}$ to at most $\frac{n_r(n_r+1)}{2}$. However, they are both greedy approaches and do not guarantee to find the globally least complex candidate model that explains the data.

\subsubsection{$l_0$ Penalized Objective Function}
An alternative approach is to penalize the number of parameters in the objective function. This can be achieved by imposing an $l_0$ penalization on the objective function (see, e.g., ~\cite{LiuLi2016}):
\begin{align}
J_{l_0}(\boldsymbol{\eta},\mathbf{r}) = J(\boldsymbol{\eta},\mathbf{r}) + \lambda\left({\dims{\eta}}+\sum_{i=1}^{\dims{r}}\left(1-\delta_{r_i}\right)\right),\qquad \mbox{with~} \delta_z=\begin{cases} 1\; \mathrm{if}\; z=0\\ 0 \;\mathrm{otherwise}\end{cases} 
\label{eq: objective function l0}
\end{align}
This $l_0$ penalization supports sparsity, i.e., reduces the model such that only a minimum number of difference parameters are used. As all models contain at least $\dims{\eta}$ parameters, only $\dims{r}$ contributes to changes in the complexity. For $\lambda = 1$, the objective function $J_{l_0}$ is the AIC divided by two. For $\lambda=\frac{1}{2}\log\left(|\mathcal{D}|\right)$, the objective function $J_{l_0}$ is the BIC divided by two. Accordingly, minimization of $J_{l_0}$ can provide the best model according to different information criteria. To directly assess the predictive power, $\lambda$ can also be determined using cross-validation.

Following \cite{RodriguezReh2013}, the objective function~\eqref{eq: objective function l0} allows for the formulation of structure inference as a mixed-integer nonlinear programming (MINLP) problem
\begin{equation}
J_{MINLP}(\boldsymbol{\eta},\mathbf{r},\mathbf{q}) = J(\boldsymbol{\eta},\mathrm{diag}(\mathbf{r}\mathbf{q}^T)) + \lambda \sum_{i=1}^{\dims{r}}q_i,
\label{eq: objective function MINLP}
\end{equation}
with real-valued $\boldsymbol{\eta} \in \mathbb{R}^{\dims{\eta}}, \mathbf{r} \in \mathbb{R}^{\dims{r}}$, and integer-valued $\mathbf{q} \in \{0,1\}^{\dims{r}}$. The optimization is done simultaneously for all parameters, $\boldsymbol{\eta}, \mathbf{r}, \mathbf{q}$. This objective function is neither differentiable, nor continuous with respect to $\mathbf{q}$. Thus, gradients with respect to the discrete parameters will not be informative for optimization. This limits the choice of optimization algorithms to derivative-free and specialized gradient-based solvers, such as the MISQP algorithm~\cite{HenriquesRoc2015}. Besides the above described and commonly used methods, further approaches, among others belief propagation~\cite{MolinelliKor2013} or iteratively reweighted least squares~\cite{DaubechiesDev2010} can be employed, under certain model assumptions. 

For the MINLP~\eqref{eq:  objective function MINLP} resulting from the $l_0$ penalized objective functions, only the comprehensive model is estimated. This, however, results in a more complex optimization problem which suffers from a high dimensional parameter space. 

\subsubsection{$l_1$ Penalized Objective Function}
To simplify the optimization problem, the $l_0$ norm is often replaced by its convex relaxation, the $l_1$ norm~\cite{Tibshirani1996}. This yields the penalized objective function
\begin{equation}
J_{l_1}(\boldsymbol{\eta},\mathbf{r}) = J(\boldsymbol{\eta},\mathbf{r}) + \lambda \sum_{i=1}^{\dims{r}}|r_i|,
\label{eq: objective function l1}
\end{equation}
with $r_i\in\mathbb{R}$, which is forced to be zero for higher $\lambda$ corresponding to the situations where the parameter $\theta_i$ has no effect. In linear regression, $l_1$ penalization is more commonly known as Lasso~\cite{Tibshirani1996}, while in signal processing it is usually referred to as Basis Pursuit~\cite{ChenDon2001}. The $l_1$ norm is continuous, but not differentiable in zero. Thus, specialized solvers have been developed which handle the non-differentiability at zero~\cite{SteiertTim2016}.

For the special case of linear regression models $J_{l_0}(\boldsymbol{\eta},\mathbf{r})$ and $J(\boldsymbol{\eta},\mathbf{r})$ are convex. 
As the $l_1$ norm is a convex relaxation of the $l_0$ norm, the resulting objective function is also convex. Thus, it can be shown that the estimated parameters $r_i$ are unique and continuous with respect to $\lambda$~\cite{EfronHas2004}. Moreover, $r_i$ can be shown to be piecewise linear with respect to $\lambda$, which allows an implementation that can efficiently compute solutions for all values of $\lambda$~\cite{EfronHas2004}. For ODE models, $J_{l_1}(\boldsymbol{\eta},\mathbf{r})$ and $J(\boldsymbol{\eta},\mathbf{r})$ will generally be non-convex and $r_i$ may be non-unique and discontinuous which is challenging for numerical methods. Thus, equation~\eqref{eq: objective function l1} is usually minimized for varying penalization strengths $\lambda$, until a reduced set of model candidates is selected. As the $l_1$ norm is an approximation of the $l_0$ norm, model selection should subsequently be performed on the reduced set of model candidates using the criteria introduced in Section~\ref{sec:model_selection} (Figure~\ref{fig:modelreduction}D).
 
The computational complexity of the $l_1$ penalization depends on the number of different penalization strengths that are used. With a higher number of tested penalizations it is more likely to obtain the globally optimal model, while a lower number of tested penalizations decreases the computational effort.

\subsubsection{Implementation and Practical Considerations}
Only few toolboxes implement methods that allow simultaneous inference of model parameters and structure. MEIGO implements the MISQP algorithm, while D2D implements the $l_1$ penalization via a modification of the fmincon routine~\cite{SteiertTim2016}.

\subsection{Model Averaging}
For large sets of candidate models and limited data, it frequently happens that not a single model is chosen by the model selection criterion. Instead, a set of models is plausible, cannot be rejected, and should be considered in the subsequent analysis. In this case, model averaging can be employed to predict the behaviour of the process (Figure~\ref{fig:modelreduction}E).

Given that a certain parameter is comparable between models, an average estimate can be derived as 
\begin{align}
\mathbb{E}[\theta_j] = \sum_m w_m \theta_{m,j}
\end{align}
with $w_m$ denoting the weight of model $M_m$ and $\theta_{m,j}$ denoting the MLE of the parameter for model $M_m$~\cite{Wassermann2000,BurnhamAnd2002}. Accordingly, uncertainties can be evaluated, e.g., the variance of the optimal values 
\begin{align}
\mathrm{var}[\theta_j] = \frac{1}{\dims M-1}\sum_m (w_m\theta_{m,j} - \mathbb{E}[\theta_j])^2.
\end{align}
The weights capture the plausibility of the model. An obvious choice is the posterior probability $p({M}_m|\mathcal{D})$. Alternatively, the Akaike weights 
\begin{align}
w_m = \frac{\exp(-\frac{1}{2}\mathrm{AIC}_m)}{\sum_{i=1}^{\dims {M}}\exp(-\frac{1}{2}\mathrm{AIC}_i)}
\end{align}
or the BIC weights 
\begin{align}
w_m = \frac{\exp(-\frac{1}{2}\mathrm{BIC}_m)}{\sum_{i=1}^{\dims {M}}\exp(-\frac{1}{2}\mathrm{BIC}_i)}.
\end{align}
can be employed. The weights for models that are not plausible are close to zero and, thus, these models do not influence the averaged model.

\section{Discussion}

This chapter provided an overview about methods for parameter inference and structure inference for ODE models of biochemical systems. For parameter inference we discussed local optimization methods and identified the number of stationary points as key determinant of computational complexity. In the context of local optimization we identified gradient-based optimization methods as suitable method as the computational complexity of determining the parameter update for optimization from the gradient of the objective function is \textit{per se} independent of the number of state variables and number of model parameters. Still, numerical optimization requires the computation of the ODE solution, which scales with the number of molecular species, and the computation of respective derivatives, which scales with the number of parameters. In both cases we discussed scaling properties of state-of-art algorithms and identified adjoint sensitivity analysis and sparse solvers as most suitable methods for large-scale problems.

We believe that key challenges to improve the scalability of parameter inference lie in the treatment of stationary points of the objective function, such as local minima and saddle points. In contrast to deep learning problems~\cite{DauphinPas2014}, the dependence of the number of stationary points on the underlying (ODE) model remains poorly understood~\cite{MannakeeRag2016,TranstrumMac2011} and should be evaluated. Local optimization methods that can account for saddle points and local minima have been developed~\cite{KanTim1987,Neumaier2004,DauphinPas2014}, but lack implementations in computational biology toolboxes and evaluations on ODE models of biochemical systems.

For structure inference, large-scale models often also give rise to a large set of different model candidates. Many model comparison criteria require parameter inference for all model candidates, which is rarely feasible if the number of model candidates is high. We discussed an $l_1$ penalization based approach that allows the simultaneous inference of model parameters and structure.
We believe that key challenges to improve the scalability of structure inference lie in the treatment of non-differentiability of the $l_1$ norm, which prohibits the application of standard gradient-based optimization algorithms. Methods such as iteratively reweighted least-squares were developed decades ago~\cite{HollandWelsch1977}, but were not adopted for ODE models.

We anticipate that, with the advent of whole cell models~\cite{KarrSan2012,BabtieStu2017,TomitaHas1999} and other large-scale models~\cite{FroehlichKes2017,BuchelRod2013}, the demand for scalable methods will drastically increase in the coming years. However, already for medium-scale models, which are much more commonplace, parameter inference and in particular structure inference can be challenging. Accordingly, there is a growing demand for novel methods with better scaling properties.

\bibliography{Database}

\begin{thebibliography}{194}
\providecommand{\natexlab}[1]{#1}
\providecommand{\url}[1]{{#1}}
\providecommand{\urlprefix}{URL }
\expandafter\ifx\csname urlstyle\endcsname\relax
  \providecommand{\doi}[1]{DOI~\discretionary{}{}{}#1}\else
  \providecommand{\doi}{DOI~\discretionary{}{}{}\begingroup
  \urlstyle{rm}\Url}\fi
\providecommand{\eprint}[2][]{\url{#2}}

\bibitem[{Klipp et~al(2005)Klipp, Herwig, Kowald, Wierling, and
  Lehrach}]{KlippBook2005}
Klipp E, Herwig R, Kowald A, Wierling C, Lehrach H (2005) Systems biology in
  practice. Wiley-VCH, Weinheim

\bibitem[{Kitano(2002{\natexlab{a}})}]{Kitano2002}
Kitano H (2002{\natexlab{a}}) Systems biology: A brief overview. Science
  295(5560):1662--1664

\bibitem[{Kitano(2002{\natexlab{b}})}]{Kitano2002b}
Kitano H (2002{\natexlab{b}}) Computational systems biology. Nature
  420(6912):206--210

\bibitem[{Adlung et~al(2017)Adlung, Kar, Wagner, She, Chakraborty, Bao,
  Lattermann, Boerries, Busch, Wuchter, Ho, Timmer, Schilling, H{\"o}fer, and
  Klingm{\"u}ller}]{AdlungKar2017}
Adlung L, Kar S, Wagner MC, She B, Chakraborty S, Bao J, Lattermann S, Boerries
  M, Busch H, Wuchter P, Ho AD, Timmer J, Schilling M, H{\"o}fer T,
  Klingm{\"u}ller U (2017) Protein abundance of {AKT} and {ERK} pathway
  components governs cell type-specific regulation of proliferation. Mol Syst
  Biol 13(1):904, \doi{10.15252/msb.20167258}

\bibitem[{Buchholz et~al(2013)Buchholz, Flossdorf, Hensel, Kretschmer,
  Weissbrich, Gr{\"a}f, Verschoor, Schiemann, H{\"o}fer, and
  Busch}]{BuchholzFlo2013}
Buchholz VR, Flossdorf M, Hensel I, Kretschmer L, Weissbrich B, Gr{\"a}f P,
  Verschoor A, Schiemann M, H{\"o}fer T, Busch DH (2013) Disparate individual
  fates compose robust {CD8+} {T} cell immunity. Science 340(6132):630--635,
  \doi{10.1126/science.1235454}

\bibitem[{Intosalmi et~al(2016)Intosalmi, Nousiainen, Ahlfors, and
  L{\"a}{\"a}hdesm{\"a}ki}]{IntosalmiNou2016}
Intosalmi J, Nousiainen K, Ahlfors H, L{\"a}{\"a}hdesm{\"a}ki H (2016)
  Data-driven mechanistic analysis method to reveal dynamically evolving
  regulatory networks. Bioinformatics 32(12):i288--i296,
  \doi{10.1093/bioinformatics/btw274}

\bibitem[{Hug et~al(2016)Hug, Schwarzfischer, Hasenauer, Marr, and
  Theis}]{HugSch2016}
Hug S, Schwarzfischer M, Hasenauer J, Marr C, Theis FJ (2016) An adaptive
  scheduling scheme for calculating {Bayes} factors with thermodynamic
  integration using {S}impson's rule. Stat Comput 26(3):663--677,
  \doi{10.1007/s11222-015-9550-0}

\bibitem[{Hross et~al(2016)Hross, Fiedler, Theis, and Hasenauer}]{HrossFie2016}
Hross S, Fiedler A, Theis FJ, Hasenauer J (2016) Quantitative comparison of
  competing {PDE} models for {Pom1p} dynamics in fission yeast. In: Findeisen
  R, Bullinger E, Balsa-Canto E, Bernaerts K (eds) Proc. 6th {IFAC} Conf.
  Found. Syst. Biol. Eng., IFAC-PapersOnLine, vol~49, pp 264--269,
  \doi{10.1016/j.ifacol.2016.12.136}

\bibitem[{Toni et~al(2012)Toni, Ozaki, Kirk, Kuroda, and Stumpf}]{ToniOza2012}
Toni T, Ozaki Yi, Kirk P, Kuroda S, Stumpf MPH (2012) Elucidating the in vivo
  phosphorylation dynamics of the {ERK} {MAP} kinase using quantitative
  proteomics data and {B}ayesian model selection. Mol Biosyst 8:1921--1929,
  \doi{10.1039/C2MB05493K}

\bibitem[{Molinelli et~al(2013)Molinelli, Korkut, Wang, Miller, Gauthier, Jing,
  Kaushik, He, Mills, Solit, Pratilas, Weigt, Braunstein, Pagnani, Zecchina,
  and Sander}]{MolinelliKor2013}
Molinelli EJ, Korkut A, Wang W, Miller ML, Gauthier NP, Jing X, Kaushik P, He
  Q, Mills G, Solit DB, Pratilas CA, Weigt M, Braunstein A, Pagnani A, Zecchina
  R, Sander C (2013) Perturbation biology: {Inferring} signaling networks in
  cellular systems. {PLoS} Comput Biol 9(12):e1003,290,
  \doi{10.1371/journal.pcbi.1003290}

\bibitem[{Schilling et~al(2009)Schilling, Maiwald, Hengl, Winter, Kreutz,
  Kolch, Lehmann, Timmer, and Klingm\"uller}]{SchillingMai2009}
Schilling M, Maiwald T, Hengl S, Winter D, Kreutz C, Kolch W, Lehmann WD,
  Timmer J, Klingm\"uller U (2009) Theoretical and experimental analysis links
  isoformspecific {ERK} signalling to cell fate decisions. Mol Syst Biol 5(334)

\bibitem[{Fey et~al(2015)Fey, Halasz, Dreidax, Kennedy, Hastings, Rauch, Munoz,
  Pilkington, Fischer, Westermann, Kolch, Kholodenko, and
  Croucher}]{FeyHal2015}
Fey D, Halasz M, Dreidax D, Kennedy SP, Hastings JF, Rauch N, Munoz AG,
  Pilkington R, Fischer M, Westermann F, Kolch W, Kholodenko BN, Croucher DR
  (2015) Signaling pathway models as biomarkers: {{Patient}-specific}
  simulations of {JNK} activity predict the survival of neuroblastoma patients.
  Sci Signal 8(408), \doi{10.1126/scisignal.aab0990}

\bibitem[{Eduati et~al(2017)Eduati, Dold{\`a}n-Martelli, Klinger, Cokelaer,
  Sieber, Kogera, Dorel, Garnett, Bl{\"u}thgen, and
  Saez-Rodriguez}]{EduatiDol2017}
Eduati F, Dold{\`a}n-Martelli V, Klinger B, Cokelaer T, Sieber A, Kogera F,
  Dorel M, Garnett MJ, Bl{\"u}thgen N, Saez-Rodriguez J (2017) Drug resistance
  mechanisms in colorectal cancer dissected with cell type-specific dynamic
  logic models. Cancer Res 77(12):3364--3375,
  \doi{10.1158/0008-5472.CAN-17-0078}

\bibitem[{Hass et~al(2017)Hass, Masson, Wohlgemuth, Paragas, Allen, Sevecka,
  Pace, Timmer, Stelling, MacBeath, Schoeberl, and Raue}]{HassMas2017}
Hass H, Masson K, Wohlgemuth S, Paragas V, Allen JE, Sevecka M, Pace E, Timmer
  J, Stelling J, MacBeath G, Schoeberl B, Raue A (2017) Predicting
  ligand-dependent tumors from multi-dimensional signaling features. npj Syst
  Biol Appl 3(1):27, \doi{10.1038/s41540-017-0030-3}

\bibitem[{Maiwald et~al(2016)Maiwald, Hass, Steiert, Vanlier, Engesser, Raue,
  Kipkeew, Bock, Kaschek, Kreutz, and Timmer}]{MaiwaldHas2016}
Maiwald T, Hass H, Steiert B, Vanlier J, Engesser R, Raue A, Kipkeew F, Bock
  HH, Kaschek D, Kreutz C, Timmer J (2016) Driving the model to its limit:
  {Profile} likelihood based model reduction. {PLoS} {ONE} 11(9),
  \doi{10.1371/journal.pone.0162366}

\bibitem[{Snowden et~al(2017)Snowden, van~der Graaf, and
  Tindall}]{SnowdenGra2017}
Snowden TJ, van~der Graaf PH, Tindall MJ (2017) Methods of model reduction for
  large-scale biological systems: {A} survey of current methods and trends. B
  Math Biol 79(7):1449--1486, \doi{10.1007/s11538-017-0277-2}

\bibitem[{Transtrum and Qiu(2016)}]{TranstrumQiu2016}
Transtrum MK, Qiu P (2016) Bridging mechanistic and phenomenological models of
  complex biological systems. {PLoS} Comput Biol 12(5):1--34,
  \doi{10.1371/journal.pcbi.1004915}

\bibitem[{Dano et~al(2006)Dano, Madsen, Schmidt, and Cedersund}]{DanoMad2006}
Dano S, Madsen MF, Schmidt H, Cedersund G (2006) Reduction of a biochemical
  model with preservation of its basic dynamic properties. FEBS Journal
  273(21):4862--4877, \doi{10.1111/j.1742-4658.2006.05485.x}

\bibitem[{Klonowski(1983)}]{Klonowski1983}
Klonowski W (1983) Simplifying principles for chemical and enzyme reaction
  kinetics. Biophys Chem 18(2):73--87, \doi{10.1016/0301-4622(83)85001-7}

\bibitem[{Hoops et~al(2006)Hoops, Sahle, Gauges, Lee, Pahle, Simus, Singhal,
  Xu, Mendes, and Kummer}]{HoopsSah2006}
Hoops S, Sahle S, Gauges R, Lee C, Pahle J, Simus N, Singhal M, Xu L, Mendes P,
  Kummer U (2006) {COPASI} -- a {CO}mplex {PA}thway {SI}mulator. Bioinformatics
  22(24):3067--3074, \doi{10.1093/bioinformatics/btl485}

\bibitem[{Raue et~al(2015)Raue, Steiert, Schelker, Kreutz, Maiwald, Hass,
  Vanlier, T{\"o}nsing, Adlung, Engesser, Mader, Heinemann, Hasenauer,
  Schilling, H{\"o}fer, Klipp, Theis, Klingm{\"u}ller, Sch{\"o}berl, and
  J.Timmer}]{RaueSte2015}
Raue A, Steiert B, Schelker M, Kreutz C, Maiwald T, Hass H, Vanlier J,
  T{\"o}nsing C, Adlung L, Engesser R, Mader W, Heinemann T, Hasenauer J,
  Schilling M, H{\"o}fer T, Klipp E, Theis FJ, Klingm{\"u}ller U, Sch{\"o}berl
  B, JTimmer (2015) {Data2Dynamics:} a modeling environment tailored to
  parameter estimation in dynamical systems. Bioinformatics 31(21):3558--3560,
  \doi{10.1093/bioinformatics/btv405}

\bibitem[{Somogyi et~al(2015)Somogyi, Bouteiller, Glazier, K{\"o}nig, Medley,
  Swat, and Sauro}]{SomogyiBou2015}
Somogyi ET, Bouteiller JM, Glazier JA, K{\"o}nig M, Medley JK, Swat MH, Sauro
  HM (2015) {libRoadRunner:} {A} high performance {SBML} simulation and
  analysis library. Bioinformatics 31(20):3315--3321,
  \doi{10.1093/bioinformatics/btv363}

\bibitem[{Santos et~al(2007)Santos, Verveer, and Bastiaens}]{SantosVer2007}
Santos SDM, Verveer PJ, Bastiaens PIH (2007) Growth factor-induced {MAPK}
  network topology shapes {Erk} response determining {PC-12} cell fate. Nat
  Cell Biol 9(3):324--330, \doi{10.1038/ncb1543}

\bibitem[{Yao et~al(2016)Yao, Pilko, and Wollman}]{YaoPil2016}
Yao J, Pilko A, Wollman R (2016) Distinct cellular states determine calcium
  signaling response. Mol Syst Biol 12(12):894, \doi{10.15252/msb.20167137}

\bibitem[{Ogilvie et~al(2017)Ogilvie, Kovachev, Wierling, Lange, and
  Lehrach}]{OgilvieKov2017}
Ogilvie LA, Kovachev A, Wierling C, Lange BMH, Lehrach H (2017) Models of
  models: {A} translational route for cancer treatment and drug development.
  Frontiers in Oncology 7:219, \doi{10.3389/fonc.2017.00219}

\bibitem[{Schillings et~al(2015)Schillings, Sunn\r{a}ker, Stelling, and
  Schwab}]{SchillingsSun2015}
Schillings C, Sunn\r{a}ker M, Stelling J, Schwab C (2015) Efficient
  characterization of parametric uncertainty of complex (bio)chemical networks.
  {PLoS} Comput Biol 11(8):e1004,457, \doi{10.1371/journal.pcbi.1004457}

\bibitem[{Babtie and Stumpf(2017)}]{BabtieStu2017}
Babtie AC, Stumpf MPH (2017) How to deal with parameters for whole-cell
  modelling. J R Soc Interface 14(133), \doi{10.1098/rsif.2017.0237}

\bibitem[{Ocone et~al(2015)Ocone, Haghverdi, Mueller, and Theis}]{OconeHag2015}
Ocone A, Haghverdi L, Mueller NS, Theis FJ (2015) Reconstructing gene
  regulatory dynamics from high-dimensional single-cell snapshot data.
  Bioinformatics 31(12):i89--i96, \doi{10.1093/bioinformatics/btv257}

\bibitem[{Kondofersky et~al(2015)Kondofersky, Fuchs, and
  Theis}]{KondoferskyFuc2015}
Kondofersky I, Fuchs C, Theis FJ (2015) Identifying latent dynamic components
  in biological systems. IET Syst Biol 9(5):193--203

\bibitem[{Akaike(1973)}]{Akaike1973}
Akaike H (1973) Information theory and an extension of the maximum likelihood
  principle. In: 2nd International Symposium on Information Theory, Tsahkadsor,
  Armenian SSR, Akademiai Kiado, vol~1, pp 267--281

\bibitem[{Schwarz(1978)}]{Schwarz1978}
Schwarz G (1978) Estimating the dimension of a model. Ann Statist
  6(2):461--464, \doi{10.1214/aos/1176344136}

\bibitem[{Steiert et~al(2016)Steiert, Timmer, and Kreutz}]{SteiertTim2016}
Steiert B, Timmer J, Kreutz C (2016) L1 regularization facilitates detection of
  cell type-specific parameters in dynamical systems. Bioinformatics
  32(17):i718--i726, \doi{10.1093/bioinformatics/btw461}

\bibitem[{Klimovskaia et~al(2016)Klimovskaia, Ganscha, and
  Claassen}]{KlimovskaiaGan2016}
Klimovskaia A, Ganscha S, Claassen M (2016) Sparse regression based structure
  learning of stochastic reaction networks from single cell snapshot time
  series. {PLoS} Comput Biol 12(12):e1005,234,
  \doi{10.1371/journal.pcbi.1005234}

\bibitem[{Loos et~al(2017)Loos, Moeller, Fr{\"o}hlich, Hucho, and
  Hasenauer}]{LoosMoe2017}
Loos C, Moeller K, Fr{\"o}hlich F, Hucho T, Hasenauer J (2017) Mechanistic
  hierarchical population model identifies latent causes of cell-to-cell
  variability. bioRxiv \doi{10.1101/171561}

\bibitem[{Hock et~al(2013)Hock, Hasenauer, and Theis}]{HockHas2013}
Hock S, Hasenauer J, Theis FJ (2013) Modeling of {2D} diffusion processes based
  on microscopy data: Parameter estimation and practical identifiability
  analysis. {BMC} Bioinf 14(Suppl 10)(S7), \doi{10.1186/1471-2105-14-S10-S7}

\bibitem[{Hross(2016)}]{HrossPhDThesis2017}
Hross S (2016) Parameter estimation and uncertainty quantification for image
  based systems biology. Ph.{D}. thesis, Technische Universit\"at M\"unchen

\bibitem[{Menshykau et~al(2013)Menshykau, Germann, Lemereux, and
  Iber}]{MenshykauGer2013}
Menshykau D, Germann P, Lemereux L, Iber D (2013) Simulating organogenesis in
  {COMSOL}: Parameter optimization for {PDE}-based models. In: Proceedings of
  COMSOL Conference, Rotterdam, Netherlands

\bibitem[{Hross and Hasenauer(2016)}]{HrossHas2016}
Hross S, Hasenauer J (2016) Analysis of {CFSE} time-series data using
  division-, age- and label-structured population models. Bioinformatics
  32(15):2321--2329, \doi{10.1093/bioinformatics/btw131}

\bibitem[{Fr\"ohlich et~al(2016)Fr\"ohlich, Thomas, Kazeroonian, Theis, Grima,
  and Hasenauer}]{FroehlichTho2016}
Fr\"ohlich F, Thomas P, Kazeroonian A, Theis FJ, Grima R, Hasenauer J (2016)
  Inference for stochastic chemical kinetics using moment equations and system
  size expansion. {PLoS} Comput Biol 12(7):e1005,030,
  \doi{10.1371/journal.pcbi.1005030}

\bibitem[{Ruess and Lygeros(2015)}]{Ruess2015}
Ruess J, Lygeros J (2015) Moment-based methods for parameter inference and
  experiment design for stochastic biochemical reaction networks. {ACM} T Math
  Softwares Model Comput S 25(2):8, \doi{10.1145/2688906}

\bibitem[{Munsky et~al(2009)Munsky, Trinh, and Khammash}]{MunskyTri2009}
Munsky B, Trinh B, Khammash M (2009) Listening to the noise: random
  fluctuations reveal gene network parameters. Mol Syst Biol 5(318),
  \doi{10.1038/msb.2009.75}

\bibitem[{Kreutz et~al(2007)Kreutz, {Bartolome Rodriguez}, Maiwald, Seidl,
  Blum, Mohr, and Timmer}]{Kreutz2007}
Kreutz C, {Bartolome Rodriguez} MM, Maiwald T, Seidl M, Blum HE, Mohr L, Timmer
  J (2007) An error model for protein quantification. Bioinformatics
  23(20):2747--2753, \doi{10.1093/bioinformatics/btm397}

\bibitem[{Maier et~al(2017)Maier, Loos, and Hasenauer}]{MaierLoo2017}
Maier C, Loos C, Hasenauer J (2017) Robust parameter estimation for dynamical
  systems from outlier-corrupted data. Bioinformatics 33(5):718--725,
  \doi{10.1093/bioinformatics/btw703}

\bibitem[{Puga et~al(2015)Puga, Krzywinski, and Altman}]{PugaKry2015}
Puga JL, Krzywinski M, Altman N (2015) Bayes' theorem. Nat Methods
  12(3):277--278, \doi{10.1038/nmeth.3335}

\bibitem[{Amestoy et~al(1996)Amestoy, Davis, and Duff}]{AmestoyDav1996}
Amestoy PR, Davis TA, Duff IS (1996) An approximate minimum degree ordering
  algorithm. {SIAM} J Matrix Anal A 17(4):886--905,
  \doi{10.1137/s0895479894278952}

\bibitem[{Dauphin et~al(2014)Dauphin, Pascanu, Gulcehre, and
  Cho}]{DauphinPas2014}
Dauphin YN, Pascanu R, Gulcehre C, Cho K (2014) Identifying and attacking the
  saddle point problem in high-dimensional non-convex optimization. In:
  Advances in Neural Information Processing Systems 26 (NIPS 2014), pp
  2933--2941

\bibitem[{Anandkumar and Ge(2016)}]{AnandkumarGe2016}
Anandkumar A, Ge R (2016) Efficient approaches for escaping higher order saddle
  points in non-convex optimization. In: Conference on Learning Theory, pp
  81--102

\bibitem[{Kirk et~al(2015)Kirk, Rolando, MacLean, and Stumpf}]{KirkRol2015}
Kirk P, Rolando DM, MacLean AL, Stumpf MP (2015) Conditional random matrix
  ensembles and the stability of dynamical systems. New J Phys 17(8):083,025,
  \doi{10.1088/1367-2630/17/8/083025}

\bibitem[{Wolpert and Macready(1997)}]{WolpertMac1997}
Wolpert DH, Macready WG (1997) No free lunch theorems for optimization. IEEE
  Trans Evol Comput 1(1):67--82, \doi{10.1109/4235.585893}

\bibitem[{Goffe et~al(1994)Goffe, Ferrier, and Rogers}]{GoffeFer1994}
Goffe WL, Ferrier GD, Rogers J (1994) Global optimization of statistical
  functions with simulated annealing. J Econometrics 60(1-2):65--99,
  \doi{10.1016/0304-4076(94)90038-8}

\bibitem[{Neumaier(2004)}]{Neumaier2004}
Neumaier A (2004) Complete search in continuous global optimization and
  constraint satisfaction. Acta Numer 13:271--369,
  \doi{10.1017/s0962492904000194}

\bibitem[{Johnson and McGeoch(1997)}]{JohnsonMcG1997}
Johnson DS, McGeoch LA (1997) The traveling salesman problem: A case study in
  local optimization. Local search in combinatorial optimization 1:215--310

\bibitem[{Hooke and Jeeves(1961)}]{HookeJee1961}
Hooke R, Jeeves TA (1961) ``{Direct Search}'' solution of numerical and
  statistical problems. J ACM 8(2):212--229, \doi{10.1145/321062.321069}

\bibitem[{Nocedal and Wright(2006)}]{NocedalWri2006}
Nocedal J, Wright S (2006) Numerical Optimization. Springer Science \& Business
  Media, \doi{10.1007/b98874}

\bibitem[{Nelder and Mead(1965)}]{NelderMea1965}
Nelder JA, Mead R (1965) A simplex method for function minimization. Comput J
  7(4):308--313, \doi{10.1093/comjnl/7.4.308}

\bibitem[{De~La~Maza and Yuret(1994)}]{DeLaMazaYur1994}
De~La~Maza M, Yuret D (1994) Dynamic hill climbing. AI expert 9:26--26

\bibitem[{Levenberg(1944)}]{Levenberg1944}
Levenberg K (1944) A method for the solution of certain non-linear problems in
  least squares. Q Appl Math 2(2):164--168, \doi{10.1090/qam/10666}

\bibitem[{Marquardt(1963)}]{Marquardt1963}
Marquardt DW (1963) An algorithm for least-squares estimation of non-linear
  parameters. {SIAM} J Appl Math 11(22):431--441, \doi{10.1137/0111030}

\bibitem[{Holland(1992)}]{Holland1992}
Holland JH (1992) Adaptation in natural and artificial systems: an introductory
  analysis with applications to biology, control, and artificial intelligence.
  MIT press

\bibitem[{Kennedy(2011)}]{Kennedy2011}
Kennedy J (2011) Particle swarm optimization. In: Encyclopedia of machine
  learning, Springer, pp 760--766

\bibitem[{Egea et~al(2007)Egea, Rodriguez-Fernandez, Banga, and
  Marti}]{Egea2007}
Egea JA, Rodriguez-Fernandez M, Banga JR, Marti R (2007) Scatter search for
  chemical and bio-process optimization. J Global Optim 37(3):481--503,
  \doi{10.1007/s10898-006-9075-3}

\bibitem[{Kirkpatrick et~al(1983)Kirkpatrick, {Gelatt~Jr}, and
  M.~P.~Vecchi}]{KirkpatrickGel1983}
Kirkpatrick S, {Gelatt~Jr} CD, M~P~Vecchi MP (1983) Optimization by simulated
  annealing. Science 220(4598):671--680, \doi{10.1126/science.220.4598.671}

\bibitem[{Kan and Timmer(1987)}]{KanTim1987}
Kan AR, Timmer GT (1987) Stochastic global optimization methods part {I}:
  {Clustering} methods. Math Program 39(1):27--56, \doi{10.1007/BF02592070}

\bibitem[{Ashyraliyev et~al(2009)Ashyraliyev, Fomekong-Nanfack, Kaandorp, and
  Blom}]{AshyraliyevFom2009}
Ashyraliyev M, Fomekong-Nanfack Y, Kaandorp JA, Blom JG (2009) Systems biology:
  {Parameter} estimation for biochemical models. FEBS Journal 276(4):886--902,
  \doi{10.1111/j.1742-4658.2008.06844.x}

\bibitem[{Fister~Jr et~al(2013)Fister~Jr, Yang, Fister, Brest, and
  Fister}]{FisterYan2013}
Fister~Jr I, Yang XS, Fister I, Brest J, Fister D (2013) A brief review of
  nature-inspired algorithms for optimization. Elektrotehniski Vestnik
  80(3):116--122

\bibitem[{Lawler and Wood(1966)}]{LawlerWoo1966}
Lawler EL, Wood DE (1966) Branch-and-bound methods: {A} survey. Oper Res
  14(4):699--719, \doi{10.1287/opre.14.4.699}

\bibitem[{T{\"o}rn and Zilinskas(1989)}]{TornZil1989}
T{\"o}rn A, Zilinskas A (1989) Global Optimization, Lecture Notes in Computer
  Science, vol 350. Springer-Verlag, Berlin

\bibitem[{Rios and Sahinidis(2013)}]{RiosSah2013}
Rios LM, Sahinidis NV (2013) Derivative-free optimization: {A} review of
  algorithms and comparison of software implementations. J Global Optim
  56(3):1247--1293, \doi{10.1007/s10898-012-9951-y}

\bibitem[{Rudolph(1994)}]{Rudolph1994}
Rudolph G (1994) Convergence analysis of canonical genetic algorithms. IEEE
  transactions on neural networks 5(1):96--101, \doi{10.1109/72.265964}

\bibitem[{Moles et~al(2003)Moles, Mendes, and Banga}]{MolesMen2003}
Moles CG, Mendes P, Banga JR (2003) Parameter estimation in biochemical
  pathways: A comparison of global optimization methods. Genome Res
  13:2467--2474, \doi{10.1101/gr.1262503}

\bibitem[{Raue et~al(2013)Raue, Schilling, Bachmann, Matteson, Schelke,
  Kaschek, Hug, Kreutz, Harms, Theis, Klingm\"uller, and Timmer}]{RaueSch2013}
Raue A, Schilling M, Bachmann J, Matteson A, Schelke M, Kaschek D, Hug S,
  Kreutz C, Harms BD, Theis FJ, Klingm\"uller U, Timmer J (2013) Lessons
  learned from quantitative dynamical modeling in systems biology. {PLoS} {ONE}
  8(9):e74,335, \doi{10.1371/journal.pone.0074335}

\bibitem[{Banga(2008)}]{Banga2008}
Banga JR (2008) Optimization in computational systems biology. {BMC} Syst Biol
  2(47), \doi{10.1186/1752-0509-2-47}

\bibitem[{Boender and Rinnooy~Kan(1987)}]{BoenderKan1987}
Boender CGE, Rinnooy~Kan AHG (1987) {Bayesian} stopping rules for multistart
  global optimization methods. Math Program 37(1):59--80,
  \doi{10.1007/BF02591684}

\bibitem[{T{\"o}rn et~al(1999)T{\"o}rn, Ali, and Viitanen}]{TornAli1999}
T{\"o}rn A, Ali MM, Viitanen S (1999) Stochastic global optimization: Problem
  classes and solution techniques. J Global Optim 14(4):437--447,
  \doi{10.1023/A:1008395408187}

\bibitem[{Fr\"ohlich et~al(2017)Fr\"ohlich, Kaltenbacher, Theis, and
  Hasenauer}]{FroehlichKal2017}
Fr\"ohlich F, Kaltenbacher B, Theis FJ, Hasenauer J (2017) Scalable parameter
  estimation for genome-scale biochemical reaction networks. {PLoS} Comput Biol
  13(1):e1005,331, \doi{10.1371/journal.pcbi.1005331}

\bibitem[{Penas et~al(2015)Penas, Gonz{\'a}lez, Egea, Banga, and
  Doallo}]{PenasGon2015}
Penas DR, Gonz{\'a}lez P, Egea JA, Banga JR, Doallo R (2015) Parallel
  metaheuristics in computational biology: {An} asynchronous cooperative
  enhanced scatter search method. Procedia Comput Sci 51:630--639,
  \doi{10.1016/j.procs.2015.05.331}

\bibitem[{Armijo(1966)}]{Armijo1966}
Armijo L (1966) Minimization of functions having {Lipschitz} continuous first
  partial derivatives. Pac J Math 16(1):1--3, \doi{10.2140/pjm.1966.16.1}

\bibitem[{Wolfe(1969)}]{Wolfe1969}
Wolfe P (1969) Convergence conditions for ascent methods. {SIAM} Rev
  11(2):226--235, \doi{10.1137/1011036}

\bibitem[{Kolda et~al(2003)Kolda, Lewis, and Torczon}]{KoldaLew2003}
Kolda TG, Lewis RM, Torczon V (2003) Optimization by direct search: {New}
  perspectives on some classical and modern methods. {SIAM} Rev 45(3):385--482,
  \doi{10.1137/s003614450242889}

\bibitem[{Lewis et~al(2000)Lewis, Torczon, and Trosset}]{LewisTor2000}
Lewis RM, Torczon V, Trosset MW (2000) Direct search methods: {Then} and now. J
  Comput Appl Math 124(1):191--207, \doi{10.1016/s0377-0427(00)00423-4}

\bibitem[{Rosenbrock(1960)}]{Rosenbrock1960}
Rosenbrock HH (1960) An automatic method for finding the greatest or least
  value of a function. Comput J 3(3):175--184, \doi{10.1093/comjnl/3.3.175}

\bibitem[{Yuan(2015)}]{Yuan2015}
Yuan Yx (2015) Recent advances in trust region algorithms. Math Program
  151(1):249--281

\bibitem[{Hartley(1961)}]{Hartley1961}
Hartley HO (1961) The modified {G}auss-{N}ewton method for the fitting of
  non-linear regression functions by least squares. Technometrics
  3(2):269--280, \doi{10.1080/00401706.1961.10489945}

\bibitem[{Nesterov and Polyak(2006)}]{NesterovPol2006}
Nesterov Y, Polyak B (2006) Cubic regularization of newton method and its
  global performance. Math Program 108(1):177--205,
  \doi{10.1007/s10107-006-0706-8}

\bibitem[{Lanczos(1950)}]{Lanczos1950}
Lanczos C (1950) An iteration method for the solution of the eigenvalue problem
  of linear differential and integral operators. United States Governm. Press
  Office Los Angeles, CA, \doi{10.1137/1.9781611971187}

\bibitem[{Nash(1984)}]{Nash1984}
Nash SG (1984) Newton-type minimization via the {Lanczos} method. SIAM Journal
  on Numerical Analysis 21(4):770--788, \doi{10.1137/0721052}

\bibitem[{Byrd et~al(1987)Byrd, Schnabel, and Shultz}]{ByrdSchnabel1987}
Byrd RH, Schnabel RB, Shultz GA (1987) A trust region algorithm for nonlinearly
  constrained optimization. {SIAM} J Numer Anal 24(5):1152--1170,
  \doi{10.1137/0724076}

\bibitem[{Sorensen(1982)}]{Sorensen1982}
Sorensen DC (1982) Newton's method with a model trust region modification.
  {SIAM} J Numer Anal 19(2):409--426, \doi{10.1137/0719026}

\bibitem[{Wild et~al(2008)Wild, Regis, and Shoemaker}]{WildRom2008}
Wild SM, Regis RG, Shoemaker CA (2008) {ORBIT:} {Optimization} by radial basis
  function interpolation in trust-regions. {SIAM} J Sci Comput
  30(6):3197--3219, \doi{10.1137/070691814}

\bibitem[{Steihaug(1983)}]{Steihaug1983}
Steihaug T (1983) The conjugate gradient method and trust regions in large
  scale optimization. SIAM Journal on Numerical Analysis 20(3):626--637,
  \doi{10.1137/0720042}

\bibitem[{Byrd et~al(1988)Byrd, Schnabel, and Shultz}]{ByrdSch1988}
Byrd RH, Schnabel RB, Shultz GA (1988) Approximate solution of the trust region
  problem by minimization over two-dimensional subspaces. Math Program
  40(1):247--263, \doi{10.1007/bf01580735}

\bibitem[{Branch et~al(1999)Branch, Coleman, and Li}]{BranchCole1999}
Branch MA, Coleman TF, Li Y (1999) A subspace, interior, and conjugate gradient
  method for large-scale bound-constrained minimization problems. {SIAM} J Sci
  Comput 21(1):1--23, \doi{10.1137/s1064827595289108}

\bibitem[{Fogel et~al(1991)Fogel, Fogel, and Atmar}]{FogelFog1991}
Fogel DB, Fogel LJ, Atmar JW (1991) Meta-evolutionary programming. In: Signals,
  systems and computers, 1991. 1991 Conference record of the twenty-fifth
  asilomar conference on, IEEE, pp 540--545, \doi{10.1109/acssc.1991.186507}

\bibitem[{Michalewicz(2013)}]{Michalewicz2013}
Michalewicz Z (2013) Genetic Algorithms + Data Structures = Evolution Programs.
  Springer Science \& Business Media

\bibitem[{Brent(2013)}]{Brent1973}
Brent RP (2013) Algorithms for minimization without derivatives. Courier
  Corporation

\bibitem[{Dembo and Steihaug(1983)}]{DemboSte1983}
Dembo RS, Steihaug T (1983) Truncated-newtono algorithms for large-scale
  unconstrained optimization. Math Program 26(2):190--212,
  \doi{10.1007/bf02592055}

\bibitem[{Solis and Wets(1981)}]{SolisWet1981}
Solis FJ, Wets RJB (1981) Minimization by random search techniques. Math Oper
  Res 6(1):19--30, \doi{10.1287/moor.6.1.19}

\bibitem[{Runarsson and Yao(2000)}]{RunarssonYao2000}
Runarsson TP, Yao X (2000) Stochastic ranking for constrained evolutionary
  optimization. {IEEE} T Evolut Comput 4(3):284--294, \doi{10.1109/4235.873238}

\bibitem[{Bellavia et~al(2004)Bellavia, Macconi, and Morini}]{BellaviaMac2004}
Bellavia S, Macconi M, Morini B (2004) {STRSCNE:} {A} scaled trust-region
  solver for constrained nonlinear equations. Comput Optim Appl 28(1):31--50,
  \doi{10.1023/b:coap.0000018878.95983.4e}

\bibitem[{Morini and Porcelli(2012)}]{MoriniPorc2012}
Morini B, Porcelli M (2012) {TRESNEI,} a matlab trust-region solver for systems
  of nonlinear equalities and inequalities. Comput Optim Appl 51(1):27--49,
  \doi{10.1007/s10589-010-9327-5}

\bibitem[{Egea et~al(2014)Egea, Henriques, Cokelaer, Villaverde, MacNamara,
  Danciu, Banga, and Saez-Rodriguez}]{EgeaHen2014}
Egea JA, Henriques D, Cokelaer T, Villaverde AF, MacNamara A, Danciu DP, Banga
  JR, Saez-Rodriguez J (2014) {MEIGO:} {An} open-source software suite based on
  metaheuristics for global optimization in systems biology and bioinformatics.
  {BMC} Bioinf 15(136), \doi{10.1186/1471-2105-15-136}

\bibitem[{Le~Digabel(2011)}]{LeDigabel2011}
Le~Digabel S (2011) Algorithm 909: {NOMAD}: Nonlinear optimization with the
  {MADS} algorithm. {ACM} T Math Software 37(4):44,
  \doi{10.1145/1916461.1916468}

\bibitem[{Kelley(1999)}]{kelley1999}
Kelley CT (1999) Iterative Methods for Optimization. SIAM

\bibitem[{W{\"a}chter and Biegler(2006)}]{WachterBie2006}
W{\"a}chter A, Biegler LT (2006) On the implementation of an interior-point
  filter line-search algorithm for large-scale nonlinear programming. Math
  Program 106(1):25--57, \doi{10.1007/s10107-004-0559-y}

\bibitem[{Ye(1989)}]{Ye1989}
Ye Y (1989) {SOLNP} users'guide. Tech. rep., Deptartment of Management
  Sciences, University of Iowa

\bibitem[{Exler et~al(2012)Exler, Lehmann, and Schittkowski}]{ExlerLeh2012}
Exler O, Lehmann T, Schittkowski K (2012) {MISQP:} a fortran subroutine of a
  trust region {SQP} algorithm for mixed-integer nonlinear programming-user's
  guide. Tech. rep., Department of Computer Science, University of Bayreuth

\bibitem[{Dennis et~al(1981)Dennis, Gay, and Welsch}]{DennisGay1981}
Dennis JE Jr, Gay DM, Welsch RE (1981) Algorithm 573: {{Nl2sol}---an} adaptive
  nonlinear least-squares algorithm {[E4]}. {ACM} T Math Software
  7(3):369--383, \doi{10.1145/355958.355966}

\bibitem[{Stapor et~al(2018)Stapor, Weindl, Ballnus, Hug, Loos, Fiedler,
  Krause, Hross, Fr\"ohlich, and Hasenauer}]{StaporWei2018}
Stapor P, Weindl D, Ballnus B, Hug S, Loos C, Fiedler A, Krause S, Hross S,
  Fr\"ohlich F, Hasenauer J (2018) {PESTO:} {Parameter} {EStimation} {TOolbox}.
  Bioinformatics 34(4):705--707, \doi{10.1093/bioinformatics/btx676}

\bibitem[{Powell(2009)}]{Powell2009}
Powell MJ (2009) The bobyqa algorithm for bound constrained optimization
  without derivatives. Tech. rep., Cambridge NA Report NA2009/06, University of
  Cambridge, Cambridge

\bibitem[{Vaz and Vicente(2009)}]{VazVic2009}
Vaz AIF, Vicente LN (2009) {PSwarm:} {A} hybrid solver for linearly constrained
  global derivative-free optimization. Optim Method Softw 24(4-5):669--685,
  \doi{10.1080/10556780902909948}

\bibitem[{Degasperi et~al(2017)Degasperi, Fey, and
  Kholodenko}]{DegasperiFey2017}
Degasperi A, Fey D, Kholodenko BN (2017) Performance of objective functions and
  optimisation procedures for parameter estimation in system biology models.
  npj Syst Biol Appl 3(1):20, \doi{10.1038/s41540-017-0023-2}

\bibitem[{Wieland(2016)}]{Wieland2016}
Wieland FG (2016) Implementation and assessment of optimization approaches for
  parameter estimation in systems biology. Tech. rep., University of Freiburg

\bibitem[{Villaverde et~al(2015)Villaverde, Henriques, Smallbone, Bongard,
  Schmid, Cicin-Sain, Crombach, Saez-Rodriguez, Mauch, Balsa-Canto, Mendes,
  Jaeger, and Banga}]{VillaverdeHen2015}
Villaverde AF, Henriques D, Smallbone K, Bongard S, Schmid J, Cicin-Sain D,
  Crombach A, Saez-Rodriguez J, Mauch K, Balsa-Canto E, Mendes P, Jaeger J,
  Banga JR (2015) {BioPreDyn}-bench: {A} suite of benchmark problems for
  dynamic modelling in systems biology. {BMC} Syst Biol 9(8),
  \doi{10.1186/s12918-015-0144-4}

\bibitem[{Kreutz(2016)}]{Kreutz2016}
Kreutz C (2016) New concepts for evaluating the performance of computational
  methods. IFAC-PapersOnLine 49(26):63--70, \doi{10.1016/j.ifacol.2016.12.104}

\bibitem[{Shamir et~al(2016)Shamir, Bar-On, Phillips, and Milo}]{ShamirBar2016}
Shamir M, Bar-On Y, Phillips R, Milo R (2016) {SnapShot:} {Timescales} in cell
  biology. Cell 164(6):1302--1302, \doi{10.1016/j.cell.2016.02.058}

\bibitem[{Hasenauer et~al(2015)Hasenauer, Jagiella, Hross, and
  Theis}]{HasenauerJag2015}
Hasenauer J, Jagiella N, Hross S, Theis FJ (2015) Data-driven modelling of
  biological multi-scale processes. Journal of Coupled Systems and Multiscale
  Dynamics 3(2):101--121, \doi{10.1166/jcsmd.2015.1069}

\bibitem[{Smallbone and Mendes(2013)}]{SmallboneMen2013}
Smallbone K, Mendes P (2013) Large-scale metabolic models: {From}
  reconstruction to differential equations. Ind Biotechnol 9(4):179--184,
  \doi{10.1089/ind.2013.0003}

\bibitem[{Resat et~al(2009)Resat, Petzold, and Pettigrew}]{ResatPet2009}
Resat H, Petzold L, Pettigrew MF (2009) Kinetic modeling of biological systems.
  Methods Mol Biol 541:311--335, \doi{10.1007/978-1-59745-243-4_14}

\bibitem[{Gonnet et~al(2012)Gonnet, Dimopoulos, Widmer, and
  Stelling}]{GonnetDim2012}
Gonnet P, Dimopoulos S, Widmer L, Stelling J (2012) A specialized {ODE}
  integrator for the efficient computation of parameter sensitivities. {BMC}
  Syst Biol 6(46), \doi{10.1186/1752-0509-6-46}

\bibitem[{Butcher(1964)}]{Butcher1964}
Butcher JC (1964) Implicit {Runge}-{Kutta} processes. Math Comp 18(85):50--64

\bibitem[{Alexander(1977)}]{Alexander1977}
Alexander R (1977) Diagonally implicit {Runge--Kutta} methods for stiff
  {O.D.E.'s}. {SIAM} J Numer Anal 14(6):1006--1021, \doi{10.1137/0714068}

\bibitem[{Rosenbrock(1963)}]{Rosenbrock1963}
Rosenbrock HH (1963) Some general implicit processes for the numerical solution
  of differential equations. Comput J 5(4):329--330,
  \doi{10.1093/comjnl/5.4.329}

\bibitem[{Zhang and Sandu(2014)}]{ZhangAdr2014}
Zhang H, Sandu A (2014) {FATODE:} {A} library for forward, adjoint, and tangent
  linear integration of {ODEs}. {SIAM} J Sci Comput 36(5):C504--C523,
  \doi{10.1137/130912335}

\bibitem[{Serban and Hindmarsh(2005)}]{SerbanHin2005}
Serban R, Hindmarsh AC (2005) {CVODES}: {T}he sensitivity-enabled {ODE} solver
  in {SUNDIALS}. In: ASME 2005 International Design Engineering Technical
  Conferences and Computers and Information in Engineering Conference, ASME, pp
  257--269, \doi{10.1115/DETC2005-85597}

\bibitem[{Hindmarsh et~al(2005)Hindmarsh, Brown, Grant, Lee, Serban, Shumaker,
  and Woodward}]{HindmarshBro2005}
Hindmarsh AC, Brown PN, Grant KE, Lee SL, Serban R, Shumaker DE, Woodward CS
  (2005) {SUNDIALS:} {S}uite of {N}onlinear and {D}ifferential/{A}lgebraic
  {E}quation {S}olvers. {ACM} T Math Software 31(3):363--396,
  \doi{10.1145/1089014.1089020}

\bibitem[{Coppersmith and Winograd(1990)}]{CoppersmithShm1990}
Coppersmith D, Winograd S (1990) Matrix multiplication via arithmetic
  progressions. J Symb Comp 9(3):251--280, \doi{10.1016/S0747-7171(08)80013-2}

\bibitem[{Thorson(1979)}]{Thorson1979}
Thorson J (1979) Gaussian elimination on a banded matrix

\bibitem[{Barabasi and Oltvai(2004)}]{BarabasiOlt2004}
Barabasi AL, Oltvai ZN (2004) Network biology: {Understanding} the cell's
  functional organization. Nat Rev Genet 5(2):101--113, \doi{10.1038/nrg1272}

\bibitem[{Davis and Palamadai~Natarajan(2010)}]{DavisPal2010}
Davis TA, Palamadai~Natarajan E (2010) Algorithm 907: {KLU}, a direct sparse
  solver for circuit simulation problems. {ACM} T Math Software 37(3):36,
  \doi{10.1145/1824801.1824814}

\bibitem[{Petzold(1983)}]{Petzold1983}
Petzold L (1983) Automatic selection of methods for solving stiff and nonstiff
  systems of ordinary differential equations. {SIAM} J Sci Stat Comp
  statistical computing 4(1):136--148, \doi{10.1137/0904010}

\bibitem[{Tangherloni et~al(2017)Tangherloni, Nobile, Besozzi, Mauri, and
  Cazzaniga}]{TangherloniNob2017}
Tangherloni A, Nobile MS, Besozzi D, Mauri G, Cazzaniga P (2017) {LASSIE:}
  {Simulating} large-scale models of biochemical systems on {GPUs}. BMC
  Bioinformatics 18(1):246, \doi{10.1186/s12859-017-1666-0}

\bibitem[{Demmel et~al(1999)Demmel, Gilbert, and Li}]{DemmelGil1999}
Demmel JW, Gilbert JR, Li XS (1999) An asynchronous parallel supernodal
  algorithm for sparse {Gaussian} elimination. {SIAM} J Matrix Anal A
  20(4):915--952, \doi{10.1137/s0895479897317685}

\bibitem[{Hucka et~al(2003)Hucka, Finney, Sauro, Bolouri, Doyle, Kitano, Arkin,
  Bornstein, Bray, Cornish-Bowden, Cuellar, Dronov, Gilles, Ginkel, Gor,
  Goryanin, Hedley, Hodgman, Hofmeyr, Hunter, Juty, Kasberger, Kremling,
  Kummer, {Le Nov\`{e}re}, Loew, Lucio, Mendes, Minch, Mjolsness, Nakayama,
  Nelson, Nielsen, Sakurada, Schaff, Shapiro, Shimizu, Spence, Stelling,
  Takahashi, Tomita, Wagner, and Wang}]{HuckaFin2003}
Hucka M, Finney A, Sauro HM, Bolouri H, Doyle JC, Kitano H, Arkin AP, Bornstein
  BJ, Bray D, Cornish-Bowden A, Cuellar AA, Dronov S, Gilles ED, Ginkel M, Gor
  V, Goryanin II, Hedley WJ, Hodgman TC, Hofmeyr JH, Hunter PJ, Juty NS,
  Kasberger JL, Kremling A, Kummer U, {Le Nov\`{e}re} N, Loew LM, Lucio D,
  Mendes P, Minch E, Mjolsness ED, Nakayama Y, Nelson MR, Nielsen PF, Sakurada
  T, Schaff JC, Shapiro BE, Shimizu TS, Spence HD, Stelling J, Takahashi K,
  Tomita M, Wagner J, Wang J (2003) The systems biology markup language
  {(SBML):} {A} medium for representation and exchange of biochemical network
  models. Bioinf 19(4):524--531, \doi{10.1093/bioinformatics/btg015}

\bibitem[{Griewank and Walther(2008)}]{GriewankWal2008}
Griewank A, Walther A (2008) Evaluating Derivatives, 2nd edn. Society for
  Industrial and Applied Mathematics, \doi{10.1137/1.9780898717761}

\bibitem[{Milne-Thompson(1933)}]{Milne1933}
Milne-Thompson L (1933) The calculus of finite differences. Macmillan

\bibitem[{Dickinson and Gelinas(1976)}]{DickinsonGel1976}
Dickinson RP, Gelinas RJ (1976) Sensitivity analysis of ordinary differential
  equation {systems---A} direct method. J Comput Phys 21(2):123--143,
  \doi{10.1016/0021-9991(76)90007-3}

\bibitem[{Kokotovic and Heller(1967)}]{KokotovicHel1967}
Kokotovic P, Heller J (1967) Direct and adjoint sensitivity equations for
  parameter optimization. IEEE T Autom Contr 12(5):609--610,
  \doi{10.1109/tac.1967.1098670}

\bibitem[{Lu et~al(2008)Lu, Muller, Machn{\'e}, and Flamm}]{LuMuller2008}
Lu J, Muller S, Machn{\'e} R, Flamm C (2008) {SBML ODE} solver library:
  Extensions for inverse analysis. In: Proc. 5th Int. W. Comp. Syst. Biol.

\bibitem[{Fujarewicz et~al(2005)Fujarewicz, Kimmel, and
  Swierniak}]{FujarewiczKimmel2005}
Fujarewicz K, Kimmel M, Swierniak A (2005) On fitting of mathematical models of
  cell signaling pathways using adjoint systems. Math Bio Eng 2(3):527--534,
  \doi{10.3934/mbe.2005.2.527}

\bibitem[{Lu et~al(2012)Lu, August, and Koeppl}]{LuAug2012}
Lu J, August E, Koeppl H (2012) Inverse problems from biomedicine: Inference of
  putative disease mechanisms and robust therapeutic strategies. J Math Biol
  67(1):143--168, \doi{10.1007/s00285-012-0523-z}

\bibitem[{Plessix(2006)}]{Plessix2006}
Plessix RE (2006) A review of the adjoint-state method for computing the
  gradient of a functional with geophysical applications. Geophys J Int
  167(2):495--503, \doi{10.1111/j.1365-246X.2006.02978.x}

\bibitem[{Balsa-Canto et~al(2001)Balsa-Canto, Banga, Alonso, and
  Vassiliadis}]{BalsaBan2001}
Balsa-Canto E, Banga JR, Alonso AA, Vassiliadis VS (2001) Dynamic optimization
  of chemical and biochemical processes using restricted second-order
  information. Comput Chem Eng 25(4):539--546,
  \doi{10.1016/s0098-1354(01)00633-0}

\bibitem[{Vassiliadis et~al(1999)Vassiliadis, Canto, and
  Banga}]{VassiliadisCan1999}
Vassiliadis VS, Canto EB, Banga JR (1999) Second-order sensitivities of general
  dynamic systems with application to optimal control problems. Chem Eng Sci
  54(17):3851--3860, \doi{10.1016/s0009-2509(98)00432-1}

\bibitem[{{\"O}zyurt and Barton(2005)}]{OzyurtBar2005}
{\"O}zyurt DB, Barton PI (2005) Cheap second order directional derivatives of
  stiff {ODE} embedded functionals. {SIAM} J Sci Comput 26(5):1725--1743,
  \doi{10.1137/030601582}

\bibitem[{Bj{\"o}rck(1996)}]{bjorck1996}
Bj{\"o}rck {\AA} (1996) Numerical Methods for Least Squares Problems. SIAM,
  \doi{10.1137/1.9781611971484}

\bibitem[{Fisher(1922)}]{Fisher1922}
Fisher RA (1922) On the mathematical foundations of theoretical statistics.
  Philos Trans R Soc London, Ser A 222:309--368, \doi{10.1098/rsta.1922.0009}

\bibitem[{Fletcher and Powell(1963)}]{FletcherPow1963}
Fletcher R, Powell MJ (1963) A rapidly convergent descent method for
  minimization. Comp J 6(2):163--168, \doi{10.1093/comjnl/6.2.163}

\bibitem[{Goldfarb(1970)}]{Goldfarb1970}
Goldfarb D (1970) A family of variable-metric methods derived by variational
  means. Math Comp 24(109):23--26, \doi{10.1090/s0025-5718-1970-0258249-6}

\bibitem[{Byrd et~al(1996)Byrd, Khalfan, and Schnabel}]{ByrdKha1996}
Byrd RH, Khalfan HF, Schnabel RB (1996) Analysis of a symmetric rank-one trust
  region method. {SIAM} J Optim 6(4):1025--1039,
  \doi{10.1137/s1052623493252985}

\bibitem[{Ramamurthy and Duffy(2017)}]{RamamurthyDuf2017}
Ramamurthy V, Duffy N (2017) {L}-{SR1}: A second order optimization method for
  deep learning, under review as a conference paper at ICLR 2017

\bibitem[{Nocedal(1980)}]{Nocedal1980}
Nocedal J (1980) Updating quasi-newton matrices with limited storage.
  Mathematics of computation 35(151):773--782, \doi{10.2307/2006193}

\bibitem[{Liu and Nocedal(1989)}]{LiuNoc1989}
Liu DC, Nocedal J (1989) On the limited memory {BFGS} method for large scale
  optimization. Math Program 45(1):503--528, \doi{10.1007/bf01589116}

\bibitem[{Andrew and Gao(2007)}]{AndrewGao2007}
Andrew G, Gao J (2007) Scalable training of l1-regularized log-linear models.
  In: Proceedings of the 24th international conference on Machine learning -
  ICML '07, ACM, pp 33--40, \doi{10.1145/1273496.1273501}

\bibitem[{Raue et~al(2009)Raue, Kreutz, Maiwald, Bachmann, Schilling,
  Klingm\"uller, and Timmer}]{Raue2009}
Raue A, Kreutz C, Maiwald T, Bachmann J, Schilling M, Klingm\"uller U, Timmer J
  (2009) Structural and practical identifiability analysis of partially
  observed dynamical models by exploiting the profile likelihood.
  Bioinformatics 25(25):1923--1929

\bibitem[{Girolami and Calderhead(2011)}]{GirolamiCal2011}
Girolami M, Calderhead B (2011) Riemann manifold {L}angevin and {H}amiltonian
  {M}onte {C}arlo methods. J R Statist Soc B 73(2):123--214,
  \doi{10.1111/j.1467-9868.2010.00765.x}

\bibitem[{Vanlier et~al(2012)Vanlier, Tiemann, Hilbers, and {van
  Riel}}]{VanlierTie2012}
Vanlier J, Tiemann CA, Hilbers PAJ, {van Riel} NAW (2012) A {Bayesian} approach
  to targeted experiment design. Bioinformatics 28(8):1136--1142,
  \doi{10.1093/bioinformatics/bts092}

\bibitem[{Blumer et~al(1987)Blumer, Ehrenfeucht, Haussler, and
  Warmuth}]{BlumerEhr1987}
Blumer A, Ehrenfeucht A, Haussler D, Warmuth MK (1987) {Occam's} razor. Inform
  Process Lett 24(6):377--380, \doi{10.1016/0020-0190(87)90114-1}

\bibitem[{Hastie et~al(2009)Hastie, Tibshirani, and Friedman}]{Hastie2009}
Hastie T, Tibshirani R, Friedman J (2009) The Elements of Statistical Learning,
  vol~2. Springer, \doi{10.1007/978-0-387-21606-5}

\bibitem[{Shibata(1980)}]{Shibata1980}
Shibata R (1980) Asymptotically efficient selection of the order of the model
  for estimating parameters of a linear process. Ann Statist pp 147--164,
  \doi{10.1214/aos/1176344897}

\bibitem[{Kass and Raftery(1995)}]{KassRaf1995}
Kass RE, Raftery AE (1995) Bayes factors. J Am Stat Assoc 90(430):773--795,
  \doi{10.2307/2291091}

\bibitem[{Wang and Sun(2014)}]{WangSun2014}
Wang M, Sun X (2014) {Bayes} factor consistency for nested linear models with a
  growing number of parameters. J Stat Plan Infer 147:95--105,
  \doi{10.1016/j.jspi.2013.11.001}

\bibitem[{Choi and Rousseau(2015)}]{ChoiRou2015}
Choi T, Rousseau J (2015) A note on {Bayes} factor consistency in partial
  linear models. J Stat Plan Infer 166:158--170,
  \doi{10.1016/j.jspi.2015.03.009}

\bibitem[{Jeffreys(1961)}]{Jeffreys1961}
Jeffreys H (1961) Theory of Probability, 3rd edn. Oxford University Press,
  Oxford

\bibitem[{Meng and Wong(1996)}]{MengWon1996}
Meng XL, Wong WH (1996) Simulating ratios of normalizing constants via a simple
  identity: a theoretical exploration. Stat Sin 6(4):831--860

\bibitem[{Skilling(2006)}]{Skilling2006}
Skilling J (2006) Nested sampling for general {Bayesian} computation. Bayesian
  Anal 1(4):833--359, \doi{10.1214/06-ba127}

\bibitem[{Burnham and Anderson(2002)}]{BurnhamAnd2002}
Burnham KP, Anderson DR (2002) Model selection and multimodel inference: {A}
  practical information-theoretic approach, 2nd edn. Springer, New York, NY

\bibitem[{Shibata(1981)}]{Shibata1981}
Shibata R (1981) An optimal selection of regression variables. Biometrika
  68(1):45--54, \doi{10.2307/2335804}

\bibitem[{Kuha(2004)}]{Kuha2004}
Kuha J (2004) {AIC} and {BIC}: Comparisons of assumptions and performance.
  Sociol Method Res 33(2):188--229, \doi{10.1177/0049124103262065}

\bibitem[{Acquah(2010)}]{Acquah2010}
Acquah HDG (2010) Comparison of {Akaike} information criterion ({AIC}) and
  {Bayesian} information criterion ({BIC}) in selection of an asymmetric price
  relationship. Journal of Development and Agricultural Economics 2(1):001--006

\bibitem[{Hurvich and Tsia(1989)}]{HurvichTsi1989}
Hurvich C, Tsia CL (1989) Regression and time series model selection in small
  samples. Biometrika 76(2):297--307, \doi{10.1093/biomet/76.2.297}

\bibitem[{Chen and Chen(2008)}]{ChenChe2008}
Chen J, Chen Z (2008) Extended {Bayesian} information criteria for model
  selection with large model spaces. Biometrika 95(3):759--771,
  \doi{10.1093/biomet/asn034}

\bibitem[{Wilks(1938)}]{Wilks1938}
Wilks SS (1938) The large-sample distribution of the likelihood ratio for
  testing composite hypotheses. Ann Math Statist 9(1):60--62,
  \doi{10.1214/aoms/1177732360}

\bibitem[{Neyman and Pearson(1992)}]{NeymanPea1992}
Neyman J, Pearson ES (1992) On the problem of the most efficient tests of
  statistical hypotheses. In: Breakthroughs in statistics, Springer, pp 73--108

\bibitem[{Gelman et~al(2014)Gelman, Hwang, and Vehtari}]{GelmanHwa2014}
Gelman A, Hwang J, Vehtari A (2014) Understanding predictive information
  criteria for {Bayesian} models. Stat Comp 24(6):997--1016,
  \doi{10.1007/s11222-013-9416-2}

\bibitem[{Arlot and Celisse(2010)}]{ArlotCel2010}
Arlot S, Celisse A (2010) A survey of cross-validation procedures for model
  selection. Statistics Surveys 4:40--79, \doi{10.1214/09-SS054}

\bibitem[{Vyshemirsky and Girolami(2008)}]{VyshemirskyGir2008}
Vyshemirsky V, Girolami M (2008) {BioBayes:} {A} software package for
  {Bayesian} inference in systems biology. Bioinformatics 24(17):1933--1934,
  \doi{10.1093/bioinformatics/btn338}

\bibitem[{Feroz et~al(2009)Feroz, Hobson, and Bridges}]{FerozHob2009}
Feroz F, Hobson M, Bridges M (2009) Multinest: an efficient and robust bayesian
  inference tool for cosmology and particle physics. Monthly Notices of the
  Royal Astronomical Society 398(4):1601--1614

\bibitem[{Thijssen et~al(2016)Thijssen, Dijkstra, Heskes, and
  Wessels}]{ThijssenDij2016}
Thijssen B, Dijkstra TM, Heskes T, Wessels LF (2016) {BCM}: toolkit for
  {B}ayesian analysis of computational models using samplers. {BMC} Syst Biol
  10(1):100

\bibitem[{Kaltenbacher and Offtermatt(2011)}]{KaltenbacherOff2011}
Kaltenbacher B, Offtermatt J (2011) A refinement and coarsening indicator
  algorithm for finding sparse solutions of inverse problems. Inverse Problems
  and Imaging 5(2):391--406, \doi{10.3934/ipi.2011.5.391}

\bibitem[{Liu and Li(2016)}]{LiuLi2016}
Liu Z, Li G (2016) Efficient regularized regression with {L0} penalty for
  variable selection and network construction. Computational and Mathematical
  Methods in Medicine 3456153, \doi{10.1155/2016/3456153}

\bibitem[{Rodriguez-Fernandez et~al(2013)Rodriguez-Fernandez, Rehberg,
  Kremling, and Banga}]{RodriguezReh2013}
Rodriguez-Fernandez M, Rehberg M, Kremling A, Banga JR (2013) Simultaneous
  model discrimination and parameter estimation in dynamic models of cellular
  systems. BMC systems biology 7(1):76, \doi{10.1186/1752-0509-7-76}

\bibitem[{Henriques et~al(2015)Henriques, M.~Saez-Rodriguez, and
  Banga}]{HenriquesRoc2015}
Henriques DR, M~Saez-Rodriguez J, Banga JR (2015) Reverse engineering of
  logic-based differential equation models using a mixed-integer dynamic
  optimization approach. Bioinformatics 31(18):2999--3007,
  \doi{10.1093/bioinformatics/btv314}

\bibitem[{Daubechies et~al(2010)Daubechies, DeVore, Fornasier, and
  G{\"u}nt{\"u}rk}]{DaubechiesDev2010}
Daubechies I, DeVore R, Fornasier M, G{\"u}nt{\"u}rk CS (2010) Iteratively
  reweighted least squares minimization for sparse recovery. Commun Pur Appl
  Math 63(1):1--38, \doi{10.1002/cpa.20303}

\bibitem[{Tibshirani(1996)}]{Tibshirani1996}
Tibshirani R (1996) Regression shrinkage and selection via the lasso. J R
  Statist Soc B 58(1):267--288, \doi{10.1111/j.1467-9868.2011.00771.x}

\bibitem[{Chen et~al(2001)Chen, Donoho, and Saunders}]{ChenDon2001}
Chen SS, Donoho DL, Saunders MA (2001) Atomic decomposition by basis pursuit.
  {SIAM} Rev 43(1):129--159, \doi{10.1137/s003614450037906x}

\bibitem[{Efron et~al(2004)Efron, Hastie, Johnstone, and
  Tibshirani}]{EfronHas2004}
Efron B, Hastie T, Johnstone I, Tibshirani R (2004) Least angle regression. Ann
  Statist 32(2):407--499, \doi{10.1214/009053604000000067}

\bibitem[{Wassermann(2000)}]{Wassermann2000}
Wassermann L (2000) {Bayesian} model selection and model averaging. J Math
  Psychol 44(1):92--107, \doi{10.1006/jmps.1999.1278}

\bibitem[{Mannakee et~al(2016)Mannakee, Ragsdale, Transtrum, and
  Gutenkunst}]{MannakeeRag2016}
Mannakee BK, Ragsdale AP, Transtrum MK, Gutenkunst RN (2016) Sloppiness and the
  Geometry of Parameter Space, Springer International Publishing, Cham, pp
  271--299. \doi{10.1007/978-3-319-21296-8_11}

\bibitem[{Transtrum et~al(2011)Transtrum, Machta, and
  Sethna}]{TranstrumMac2011}
Transtrum MK, Machta BB, Sethna JP (2011) Geometry of nonlinear least squares
  with applications to sloppy models and optimization. Phys Rev E 83:036,701,
  \doi{10.1103/PhysRevE.83.036701}

\bibitem[{Holland and Welsch(1977)}]{HollandWelsch1977}
Holland PW, Welsch RE (1977) Robust regression using iteratively reweighted
  least-squares. Commun Stat Theory 6(9):813--827,
  \doi{10.1080/03610927708827533}

\bibitem[{Karr et~al(2012)Karr, Sanghvi, Macklin, Gutschow, Jacobs,
  {Bolival~Jr}, Assad-Garcia, Glass, and Covert}]{KarrSan2012}
Karr JR, Sanghvi JC, Macklin DN, Gutschow MV, Jacobs JM, {Bolival~Jr} B,
  Assad-Garcia N, Glass JI, Covert MW (2012) A whole-cell computational model
  predicts phenotype from genotype. Cell 150(2):389--401,
  \doi{10.1016/j.cell.2012.05.044}

\bibitem[{Tomita et~al(1999)Tomita, Hashimoto, Takahashi, Shimizu, Matsuzaki,
  Miyoshi, Saito, Tanida, Yugi, Venter, and {Hutchision~III}}]{TomitaHas1999}
Tomita M, Hashimoto K, Takahashi K, Shimizu TS, Matsuzaki Y, Miyoshi F, Saito
  K, Tanida S, Yugi K, Venter JC, {Hutchision~III} CA (1999) E-{CELL:}
  {Software} environment for whole-cell simulation. Bioinformatics
  15(1):72--84, \doi{10.1093/bioinformatics/15.1.72}

\bibitem[{Fr\"ohlich et~al(2017)Fr\"ohlich, Kessler, Weindl, Shadrin,
  Schmiester, Hache, Muradyan, Schuette, Lim, Heinig, Theis, Lehrach, Wierling,
  Lange, and Hasenauer}]{FroehlichKes2017}
Fr\"ohlich F, Kessler T, Weindl D, Shadrin A, Schmiester L, Hache H, Muradyan
  A, Schuette M, Lim JH, Heinig M, Theis F, Lehrach H, Wierling C, Lange B,
  Hasenauer J (2017) Efficient parameterization of large-scale mechanistic
  models enables drug response prediction for cancer cell lines. bioRxiv
  \doi{10.1101/174094}

\bibitem[{B\"uchel et~al(2013)B\"uchel, Rodriguez, Swainston, Wrzodek,
  Czauderna, Keller, Mittag, Schubert, Glont, Golebiewski, {van}~Iersel,
  Keating, Rall, Wybrow, Hermjakob, Hucka, Kell, M\"uller, Mendes, Zell,
  Chaouiya, Saez-Rodriguez, Schreiber, Laibe, Dr\"ager, and
  Nov\'ere}]{BuchelRod2013}
B\"uchel F, Rodriguez N, Swainston N, Wrzodek C, Czauderna T, Keller R, Mittag
  F, Schubert M, Glont M, Golebiewski M, {van}~Iersel M, Keating S, Rall M,
  Wybrow M, Hermjakob H, Hucka M, Kell DB, M\"uller W, Mendes P, Zell A,
  Chaouiya C, Saez-Rodriguez J, Schreiber F, Laibe C, Dr\"ager A, Nov\'ere NL
  (2013) {Path2Models:} {{Large}-scale} generation of computational models from
  biochemical pathway maps. {BMC} Syst Biol 7(116),
  \doi{10.1186/1752-0509-7-116}

\end{thebibliography}

\end{document}